\documentclass[pre,aps,amsmath,amssymb,showpacs]{revtex4}
\usepackage{graphicx,epsfig}
\usepackage{wasysym}

\usepackage{amsmath,graphicx}
\usepackage{epsfig}
\usepackage{bm}
\usepackage{color}
\usepackage{amssymb}

\begin{document}

\title{Force spectroscopy of polymer desorption: Theory and Molecular Dynamics
simulation
}
\author{Jaros{\l}aw Paturej\textit{$^{a,b}$}, Johan L.A. Dubbeldam{$^{c}$}, Vakhtang G. Rostiashvili{$^{d}$}, Andrey Milchev\textit{$^{d,e}$} and
Thomas A. Vilgis\textit{$^{d}$}}
\affiliation{\textit{$^{a}$~Department of Chemistry,University of North Carolina, Chapel Hill, NC
27599, USA}\\
\textit{$^{b}$~Institute of Physics, University of Szczecin, Wielkopolska 15,
70451 Szczecin, Poland}\\
\textit{$^{\ast}$ Corresponding author: paturej@live.unc.edu}\\
\textit{$^{c}$~Delft University of Technology 2628CD Delft, The Netherlands}\\
\textit{$^{d}$~Max Planck Institute for Polymer Research, 10 Ackermannweg,
55128 Mainz, Germany}\\
\textit{$^{e}$~Institute of Physical Chemistry, Bulgarian Academy of
Sciences, 1113 Sofia, Bulgaria}
}

\begin{abstract}
Forced detachment of a single polymer chain, strongly-adsorbed on a solid
substrate, is investigated by two complementary methods: a coarse-grained
analytical dynamical model, based on the Onsager stochastic equation, and
Molecular Dynamics (MD) simulations with Langevin thermostat. The suggested
approach makes it possible to go beyond the limitations of the conventional
Bell-Evans model. We observe a series of characteristic force spikes when the
pulling force is measured against the cantilever displacement  during
detachment at constant velocity $v_c$ (displacement control mode) and find that
the average magnitude of
this force increases as $v_c$ grows. The probability  distributions of the
pulling force and the end-monomer distance from the surface at the moment of
final detachment are investigated for different adsorption energy $\epsilon$ and
pulling velocity $v_c$. Our extensive MD-simulations validate and support the
main theoretical findings. Moreover, the simulation reveals a novel behavior:
for a strong-friction and massive cantilever the force spikes pattern is smeared
out at large $v_c$. As a challenging task for experimental bio-polymers
sequencing in future we suggest the fabrication of stiff, super-light,
nanometer-sized AFM probe.
\end{abstract}


\maketitle

\section{Introduction}

In recent years single-molecule pulling techniques based on the use of laser
optical tweezers (LOT) or atomic force microscope (AFM) have gained prominence
as a versatile tool in the studies of non-covalent bonds and self-associating
bio-molecular systems
\cite{Ritort,Franco,Hugel,Butt,Gao,Janshof,Carrion,Fisher,Gaub}. The
latter could be exemplified by the base-pair binding in DNA as well as by
ligand-receptor interactions in proteins and  has been studied recently by means
of Brownian dynamic simulations and the master equation approach
\cite{Alexander_1,Alexander_2}.  The LOT and AFM methods are commonly used to
manipulate and exert mechanical forces on individual molecules. In LOT
experiments, a micron-sized polystyrene or silica bead is trapped in the focus
of the laser beam by exerting forces in the range $0.1 - 100\: pN$. Typically,
AFM (which covers forces interval in $20\: pN - 10\: nN$ range) is ideal for
investigations of relatively strong inter- or intramolecular interactions which
are involved in pulling experiments in biopolymers such as polysaccharides,
proteins and nucleic acids. On the other hand, due to the relatively small
signal-to-noise ratio, the AFM experiments have limitations with regard to the
mechanochemistry of weak interactions in the lower piconewton regime.

The method of dynamic force spectroscopy (DFS) is used to probe the
force-extension relationship, rupture force distribution, and the force vs
loading rate dependence for single-molecule bonds or for more complicated
multiply-bonded attachments. Historically, the first theoretical interpretation
of DFS has been suggested in the context of single cell adhesion by Bell
\cite{Bell} and developed by Evans \cite{Evans_1,Evans_2,Evans_3}. The
consideration has been based on the semi-phenomenological Arrhenius relation
which describes surface detachment under time-dependent pulling force, $f = r_l
t$, with $r_l$ being the loading rate. It was also assumed that the effective
activation energy, $E_{b} (f)$, may be approximated by a linear function of the
force, i.e.,  $E_{b} (f) = E_{b}^{(0)} - x_{\beta} f$. Here $x_{\beta}$ is the
distance between the bonded state and the transition state where the activation
barrier is located. The resulting Bell-Evans (BE) equation then gives the mean
detachment force as a function of temperature $T$ and loading rate $r_l$, i.e., $f
= \frac{k_BT}{x_{\beta}} \ln (\frac{r x_{\beta}}{k_BT \kappa_0})$, where
$\kappa_0$ is the desorption rate in the absence of applied pulling force.

As one can see from this BE-equation, the simple surmounting of  BE-activation
barrier results in a linear dependence of detachment force on the logarithm of
loading rate, provided one uses the applied force as a governing parameter in
the detachment process (i.e., working in an {\em isotensional} ensemble when $f$
is controlled and the distance $D$ from the substrate to the clamped end-monomer
of the polymer chain fluctuates). For multiply-bonded attachments the
interpretation problem based on this equation becomes more complicated since a
non-linear $f-\ln r_l$ relationship is observed \cite{Merkel}. In this case
chain detachment involves passages over a cascade of activation barriers. For
example, Merkel et al. \cite{Merkel} suggested that the net rate of detachments
can be approximated by a reciprocal sum of characteristic times, corresponding
to jumps over the single barriers. In particular, regarding the detachment of
biotin-streptavidin single bonds, it was suggested that two consecutive
barriers might be responsible for the desorption process.

A simple example of multiply-bonded bio-assembly is presented by a
singe-stranded DNA (ssDNA) macromolecule, strongly adsorbed on graphite
substrate. The forced-induced desorption (or peeling) of this biopolymer has
been studied analytically and by means of Brownian dynamics (BD) simulation by
Jagota et al. \cite{Jagota_1,Jagota_2,Jagota_3,Jagota_4}. In ref.
\cite{Jagota_1} the equilibrium statistical thermodynamics of ssDNA
forced-induced desorption under force control (FC) and displacement control (DC)
has been investigated. In the latter case one works in an {\em isometric}
ensemble where $D$ is controlled and $f$ fluctuates. It has been demonstrated
that the force response under DC exhibits a series of spikes which carry
information about the underlying base sequence of ssDNA. The Brownian dynamics
(BD) simulations \cite{Jagota_2} confirmed the existence of such force spikes in
the force-displacement curves under DC.

The nonequilibrium theory of forced desorption has been developed by Kreuzer et
al. \cite{Kreuzer_1,Kreuzer_2,Kreuzer_3} on the basis of Master Equation
approach for the cases of constant velocity and force-ramp modes in an
AFM-experiment. The authors assumed that individual monomers detachments
represent a fast process as compared to the removal of all monomers. This
justifies a two-state model where all monomers either remain on the substrate or
leave it abruptly. The corresponding transition rates (which constitute  a
necessary input in the Master Equation approach) must satisfy detailed balance.
As a result of the Master Equation solution, the authors obtained a probability
distribution of detachment heights (i.e., distances between the cantilever tip
and the substrate) as well as an average detachment height as a function of the
pulling velocity.

Irrespective of all these  efforts, a detailed theoretical interpretation  of
the dynamic force spectroscopy experiments is still missing. For example, in
terms of Kramers reaction-rate theory \cite{Hanggi} the Arrhenius-like BE -
model holds only when the effective activation energy $E_b (f) \gg k_BT$. On the
other hand, it is clear that for large forces (which we experience in AFM), the
case when $E_b (f) \approx k_BT$ occurs fairly often. In this common case the
general approach, based on the BE-model, becomes questionable. Besides, it can
be shown \cite{Hanke}, that the activation energy vs force dependence,
$E_{b}(f)$, is itself a {\em nonlinear} function, so that the conventional BE -
model, based on the linear approximation, $E_{b} (f) \approx E_{b}^{(0)} -
x_{\beta} f$, should be limited to small forces. Moreover, the Arrhenius - like
relationship for the detachment rate, which was used in the BE - model, is a
consequence of a saddle-point approximation for the stationary solution of
Fokker- Planck equation \cite{Hanggi}. This contradicts the typical loading
regimes, used in experiments, where applied force or distance grow linearly with
time.

The present paper is devoted to the theoretical investigation of a single
molecule desorption dynamics and aimed at interpretation of AFM - or LOT - based
dynamic force spectroscopy in the DC constant-velocity mode. The organization of
the paper is as follows: in Sec. \ref{Second_Section} we give the equilibrium
theory of detachment for the case of strong polymer adsorption. The mean  force
(measured at the cantilever tip) versus displacement diagram is discussed in
detail. In particular, the characteristic force-``spikes'' structure (which was
first discussed in Ref. \cite{Jagota_1,Jagota_2}) can be clearly seen. In Sec.
\ref{sec_dynamics} we give a dynamical version of the detachment process. Our
approach rests on construction of general free energy functions, depending on
coarse-grained variables, which govern the non-linear response and structural
bonding changes in presence of external forces. The corresponding
free-energy-based stochastic equations (known as Onsager equations \cite{Puri})
are derived and solved numerically. This solution makes it possible to provide
not only force-displacement diagrams and the ensuing dependence on cantilever
displacement velocity $v_c$ but also the detachment force probability
distribution function (PDF). In Sec. \ref{sec_MD} the main theoretical results
are then checked against extensive Molecular Dynamics (MD) simulation. A brief
discussion of results is offered in Sec. \ref{sec_results}.

\section{Equilibrium theory at the strong adsorption case}
\label{Second_Section}

\begin{figure}[ht]
\begin{center}
\includegraphics[width =7.0cm]{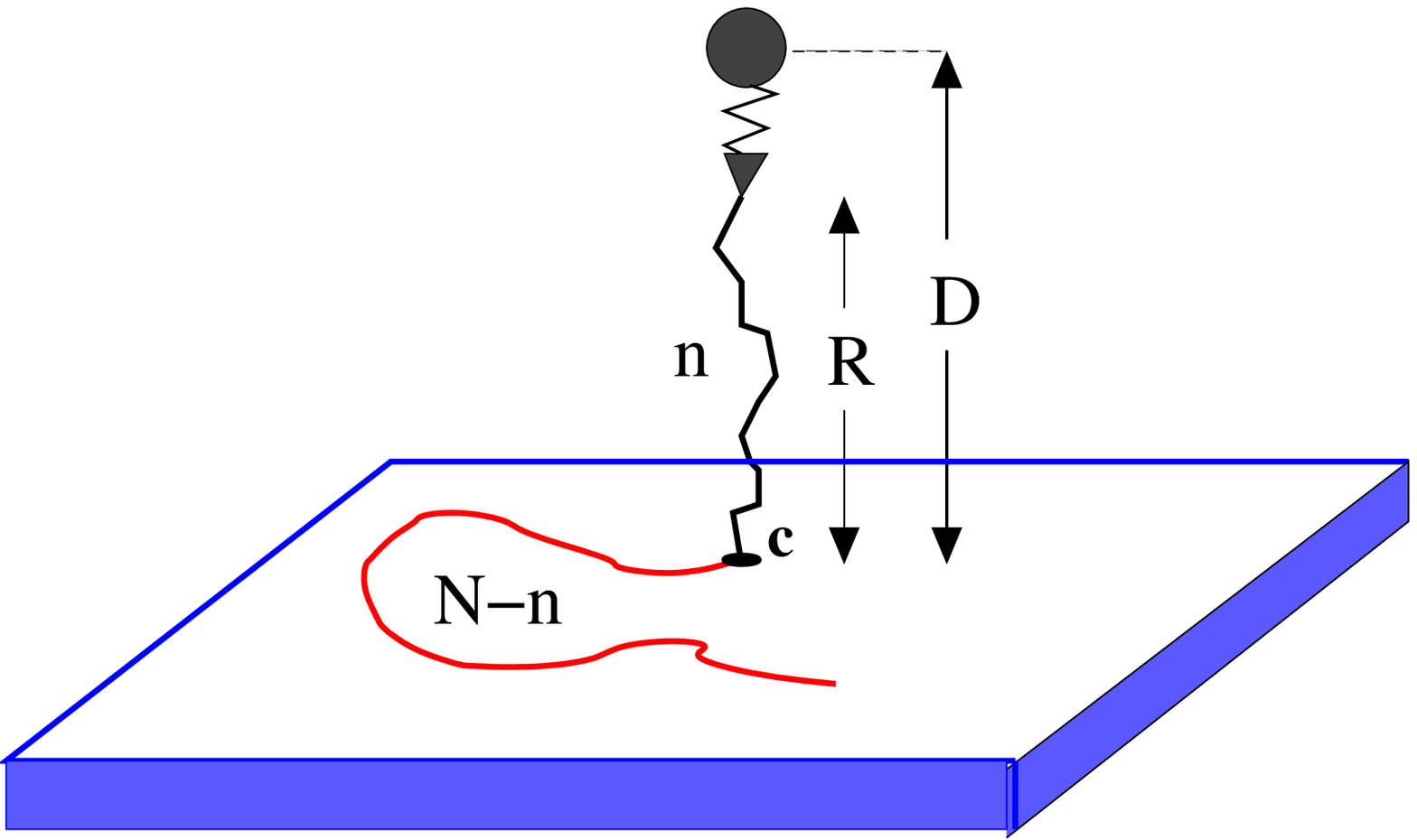}
\hspace{1cm}
\includegraphics[width =8.0cm]{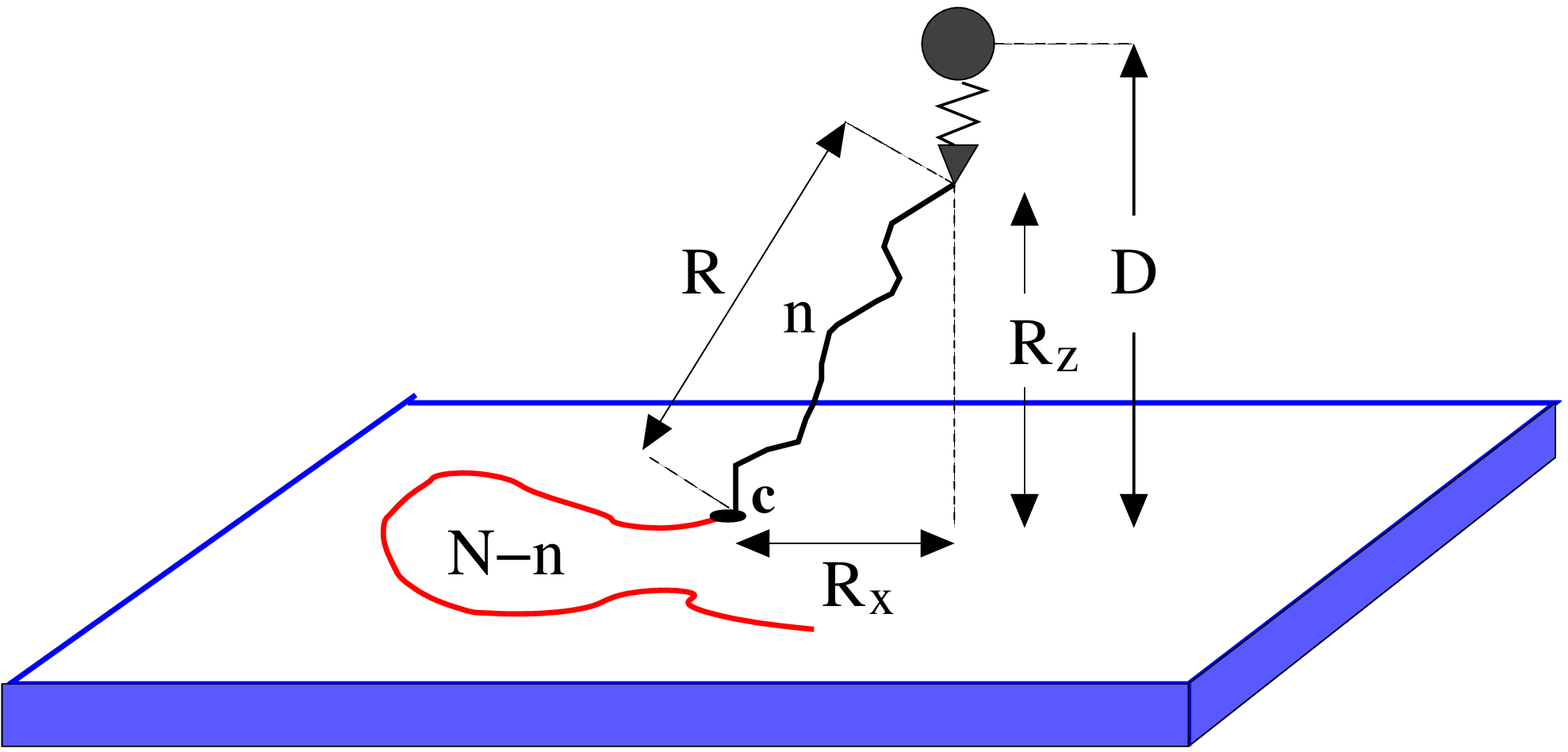}
\caption{Principal scheme of a single molecule forced desorption experiment
based on the AFM: (left panel) There is no friction between the adsorbed
portion and substrate. (right panel) Strong friction case.}
\label{Geometry}
\end{center}
\end{figure}

Recently we suggested a theory of the force-induced polymer desorption (for
relatively weak adsorption energy) in the isotensional
\cite{Bhattacharya_1,Bhattacharya_2} and isometric \cite{Bhattacharya_3}
equilibrium ensembles supported by extensive Monte Carlo (MC) simulations. In
the former case, the fraction of adsorbed monomers
changes abruptly (undergoes a jump) when one varies the adsorption energy or
the external pulling force. In the second case, the order parameter varies
steadily with changing height of the AFM-tip, even though the phase transition
is still of first order. The total phase diagram in terms of adsorption energy
- pulling force, or, adsorption energy - end-monomer height, has been discussed
theoretically and in terms of MC-simulations.

On the other hand,  the AFM experiments deal with relatively strong forces
($20\:pN - 10\:nN$ \cite{Ritort}) so that in the case of a single molecule
desorption experiment only a really strong adsorption energy is essential. This
limit has been discussed in the recent papers by Jagota et al.
\cite{Jagota_1,Jagota_2,Jagota_3,Jagota_4} and Kreuzer ey al.
\cite{Kreuzer_1,Kreuzer_2,Kreuzer_3}. Here we consider this problem in a
slightly more general form. In so doing we distinguish between two different
models: with frictionless- and strong-friction substrates, as indicated in Fig.
\ref{Geometry}.

\subsection{Frictionless substrate}

This case has been considered in Refs.
\cite{Jagota_1,Jagota_2,Kreuzer_1,Kreuzer_2,Kreuzer_3} and is based on the
assumption that the force resisting sliding is sufficiently small, i.e., the
cantilever tip and the contact point $c$ are both placed along the same $z$-axis
(see Fig. \ref{Geometry} (left panel)). The total partition function for a fixed
cantilever distance $D$, i.e.,  $\Xi_{\rm tot} (D)$, is a product of partition
functions of the adsorbed part , $\Xi_{\rm ads} (n)$ , of the desorbed portion
(a stretched polymer portion), $\Xi_{\rm pol} (n, R)$,  and of the cantilever
itself, $\Xi_{\rm can} (D-R)$, where $n$ is the number of {\em desorbed} polymer
segments, and $R$ denotes the distance between the clamped end of this desorbed
portion and the substrate. As a result,
\begin{eqnarray}
 \Xi_{\rm tot} (D) = \sum_{n=0}^{N} \: \Xi_{\rm ads}
(n) \: \int\limits_{0}^{b n} \: d R \; \Xi_{\rm pol} (n, R)\: \theta (D - R)
\Xi_{\rm can} (D-R),
\label{Combination}
\end{eqnarray}
where the integration interval, $ 0 < R < b n$, and the step-function, $\theta(
D - R)$, imply that restrictions, $R < b n$ and $ R < D$, should be applied
simultaneously. In this representation $D$ is the control variable (which is
monitored by the corresponding AFM operating mode) whereas $n$ and $R$ are
coarse-grained dynamic variables which should be integrated (in our case, an
integral over $R$, and summation over $n$) out. Moreover, if we introduce the
function
\begin{eqnarray}
 \min (b n, D) = \begin{cases}
                  b n &\mbox{for} \qquad bn < D\\
                  D &\mbox{for} \qquad D < b n,
 \label{Min}                \end{cases}
\end{eqnarray}
then Eq. (\ref{Combination}) can be rewritten as
\begin{eqnarray}
 \Xi_{\rm tot} (D) = \sum_{n=0}^{N} \: \Xi_{\rm ads}
(n) \: \int\limits_{0}^{\min (b n, D)} \: d R \; \Xi_{\rm pol} (n, R)\:
\Xi_{\rm can} (D-R)
\label{Combination_1}
\end{eqnarray}

In the strong adsorption regime, $\Xi_{\rm ads} (n)$ attains a simple form
 \begin{eqnarray}
  \Xi_{\rm ads} (n) = \exp \left[ \epsilon (N - n)\right],
\label{Adsorption_Z}
 \end{eqnarray}
where the dimensionless adsorption energy $\epsilon = \varepsilon/k_BT$. The
cantilever manifests itself as a harmonic spring with a spring constant
$k_c$, i.e., the corresponding partition function reads
 \begin{eqnarray}
  \Xi_{\rm can} (D-R) = \exp \left[ - \dfrac{k_c}{2k_BT} (D - R)^2 \right]
 \label{Cantilever}
\end{eqnarray}

Finally, we derive the partition function of the desorbed part of the polymer
as function of the dynamic variables $n$ and $R$, based of the {\em Freely
Jointed Bond Vector} (FJBV) model \cite{Lai,Schurr}. The corresponding Gibbs
free energy (i.e., the free energy in the isotensional-ensemble) is
\begin{eqnarray}
 G_{\rm pol} (n, {\widetilde f}) = - n k_BT \: \ln \left[ \dfrac{\sinh
{\widetilde
f}}{{\widetilde f}}\right],
\label{Gibbs}
\end{eqnarray}
where the dimensionless force ${\widetilde f}  \stackrel{\rm def}{=} b f/k_BT$.
The corresponding distance $R = - \partial G_{\rm pol} (n, {\widetilde f})/
\partial {\widetilde f}$.
\begin{eqnarray}
 R = - \dfrac{\partial G_{\rm pol}}{\partial f} = n b {\cal L} ({\widetilde f}),
\label{Langevin}
\end{eqnarray}
where the so called Langevin function ${\cal L} ({\widetilde f}) \equiv \coth
({\widetilde f}) - 1/{\widetilde f}$ has been used. In the isometric-ensemble,
the proper thermodynamic potential is the Helmholtz free energy, $F_{\rm pol}
(n,R)$, which is related to $G_{\rm pol} (n, {\widetilde f})$ by Legendre
transformation,
\begin{eqnarray}
 F_{\rm pol} (n,R) = G_{\rm pol} (n, {\widetilde f}) + f \: R,
\label{Legander}
\end{eqnarray}
where $f = \partial F_{\rm p} (n,R) /\partial R$. Taking the Gibbs free energy,
Eq. (\ref{Gibbs}), into account and the relation Eq. (\ref{Langevin}) for the
Helmholtz free energy, we have
\begin{eqnarray}
 F_{\rm pol} (n,R) &=& G_{\rm pol} (n, {\widetilde f}) + f \: R \nonumber\\
 &=& - n \: k_BT \: \ln \left[ \dfrac{\sinh {\widetilde
f}}{{\widetilde f}}\right]  + {\widetilde f} \: k_BT \: n \: {\cal L}
({\widetilde
f})\nonumber\\
&=& - n \: k_BT \; \left\lbrace
\ln \left[ \dfrac{\sinh {\widetilde f}}{{\widetilde f}}\right] + 1 - {\widetilde
f} \: \coth ({\widetilde f} ) \right\rbrace  \equiv  - n \: k_BT \; {\cal G}
({\widetilde f}),
\label{Helmholtz}
\end{eqnarray}
where the function ${\cal G} (x) \equiv  \ln [ \sinh (x)/x] + 1 - x \coth (x)$.
As a result, Eq. (\ref{Helmholtz}) along with Eq. (\ref{Langevin})
parametrically define $F_{\rm p} (n,R)$, and the corresponding partition
function
\begin{eqnarray}
 \Xi_{\rm pol} (n, R) = \exp\left[ n {\cal G} ({\widetilde f})\right]
\label{Polymer_Z}
\end{eqnarray}
as function of $n$ and $R$.

By making use of Eqs. (\ref{Adsorption_Z}), (\ref{Cantilever}),
(\ref{Polymer_Z}),
the total partition function given by Eq. (\ref{Combination}) reads
\begin{eqnarray}
\Xi_{\rm tot} (D) =  \sum_{n=0}^{N} \: \int\limits_{0}^{\min (n b, D)} \: d R\;
\exp\left[
 \epsilon (N - n)\right] \: \exp \left[ n \: {\cal G} ({\widetilde f}) \right]
\: \exp \left[ - \dfrac{k_c}{2 k_BT} (D - R)^2 \right].
\label{Total_Z_Final}
\end{eqnarray}
In Eq. (\ref{Total_Z_Final}) the force ${\widetilde f}$  should be expressed in
terms of $R/b n$ as follows: ${\widetilde f} = {\cal L}^{-1} (R/b n)$, where
${\cal L}^{-1} (x)$ denotes the inverse Langevin function.
The corresponding effective free energy function in terms of $n$ and $R$ reads
\begin{eqnarray}
 {\cal F} (n, R) = - k_BT \epsilon (N - n) - k_BT n {\cal G} ({\widetilde f}) +
\dfrac{k_c}{2 } \left(D - R \right)^2
\label{Free_Energy_Frictionless}
\end{eqnarray}

In the limit of a very stiff cantilever, $k_c b^2/k_BT \gg 1$, the cantilever
partition function approaches a $\delta$-function \cite{Kreuzer_1}:
\begin{eqnarray}
  \Xi_{\rm can} (D-R) = \exp \left[ - \dfrac{k_c}{2 k_BT} (D - R)^2 \right]
\rightarrow  ( 2\pi k_BT/k_c )^{1/2}  \: \delta (D - R),
\label{Delta}
\end{eqnarray}
and Eq. (\ref{Total_Z_Final}) takes the form
\begin{eqnarray}
 \Xi_{\rm tot} (D) = \sum_{n=0}^{N} \: \exp\left[
 \epsilon (N - n)\right] \: \exp \left[ n \: {\cal G} ({\widetilde f}) \right]
\: \theta (n b - D),
\label{Total_Z_Simple}
\end{eqnarray}
where ${\widetilde f} = {\cal L}^{-1} (D/b n)$ and the step-function $\theta
(b n - D)$ ensures that the condition $b n > D$ holds. It is this very stiff
cantilever limit that was considered in ref. \cite{Jagota_1,Jagota_2}.

For the isometric ensemble, i.e., in the $D$-ensemble, the average
force $\langle f_{z} \rangle$, measured by AFM-experiment, is given by
 \begin{eqnarray}
  \langle f_{z} \rangle &=& - k_BT \: \dfrac{\partial}{\partial D} \: \ln
\Xi_{\rm tot} (D)\nonumber\\
&=& \dfrac{k_c}{ \Xi_{\rm tot} (D) } \:
  \sum_{n=0}^{N} \:
\exp\left[\epsilon (N - n)\right] \: \int\limits_{0}^{\min (b n, D)} \: d R \;
(D - R) \: \exp \left[ n \: {\cal G} ({\widetilde f})
\right]\: \exp \left[ - \dfrac{k_c}{2 k_BT} (D - R)^2 \right],
 \label{Average_Force_1}
\end{eqnarray}
where $\Xi_{\rm tot} (D)$ is given by Eq. (\ref{Total_Z_Final}).

\begin{figure}[ht]
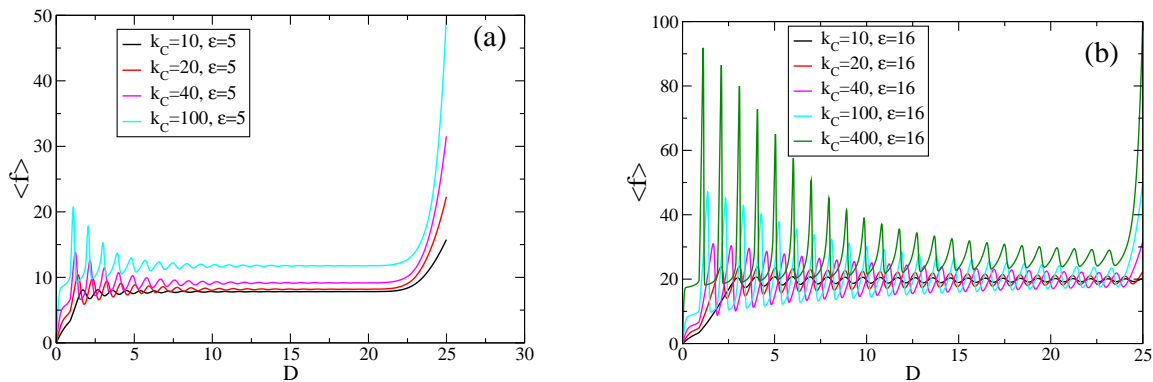

\begin{center}
\vspace{0.7cm}
\includegraphics[width =7.0cm]{ep5kcgraphR.eps}
\hspace{1cm}
\includegraphics[width =7.0cm]{ep16kcgraphR.eps}

\caption{The equilibrium force-displacement diagrams calculated according to Eq.
(\ref{Average_Force_1}). The sawtooth structure becomes more pronounced with
increasing adsorption energy $\epsilon$ and spring constant $k_c$.}
\label{Equilibrium}
\end{center}
\end{figure}

The numerical results, which follow from Eq. (\ref{Average_Force_1}), are shown
in Fig. \ref{Equilibrium}. One can immediately see the "sawtooth``-, or
force-spikes structure on the force-displacement diagram as it was also found by
Jagota et al. \cite{Jagota_1} in the limit of very stiff cantilever.
Physically, spikes correspond to the reversible transitions $n\rightleftarrows
n + 1$, during which  the release of polymer stretching energy is balanced by
the adsorption energy. The corresponding thermodynamic
condition reads ${\cal F}(n, R) = {\cal F}(n + 1, R)$. This condition
also leads to the spikes amplitude law $f_{\rm amp} \propto
\exp (\epsilon/n)$  \cite{Jagota_1}, i.e. the spikes
amplitude gradually decreases in the process of chain detachment (i.e., with
growing $n$).

\begin{figure}[ht]
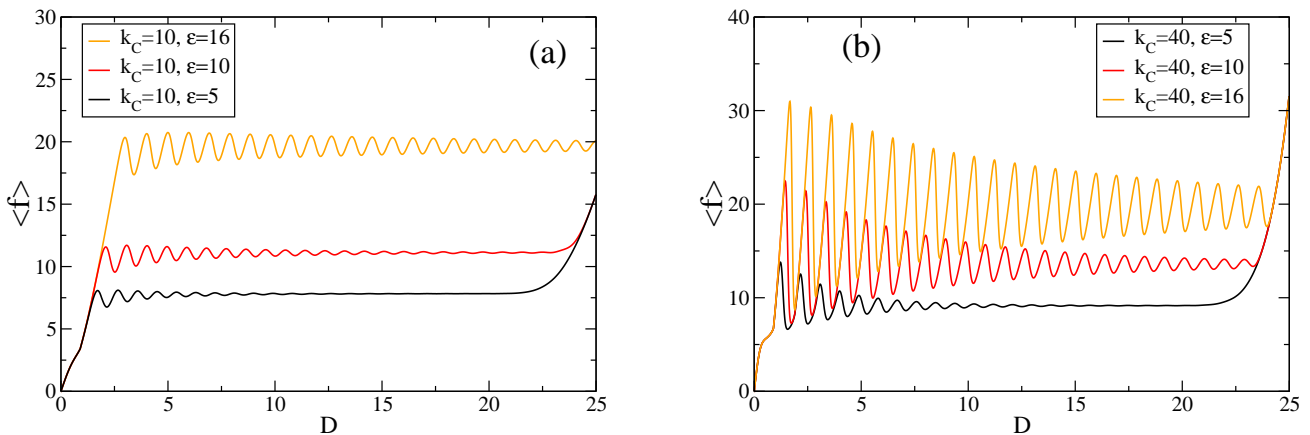

\begin{center}
\vspace{0.7cm}
\includegraphics[width =8.0cm]{kc10epgraphR.eps}
\hspace{1cm}
\includegraphics[width =8.0cm]{kc40epgraphR.eps}

\caption{The equilibrium force-displacement diagrams calculated according to Eq.
~(\ref{Average_Force_1}). The same as in Fig.~\ref{Equilibrium} but for fixed
$k_c$  and different $\epsilon$. }
\label{Equilibrium_1}
\end{center}
\end{figure}

This
structure is more pronounced at larger adsorption energy $\epsilon$ and
cantilever spring constant $k_c$.  Thus, while the force oscillates, its mean
value remains nearly constant in a broad interval of distances $D$, exhibiting a
kind of plateau.
Complementary information (for fixed $k_c$ at different values of $\epsilon$)
is given on Fig. \ref{Equilibrium_1}. One can verify that the plateau height is
mainly determined by $\epsilon$ whereas the spikes amplitude is dictated by the
cantilever spring constant $k_c$.

\subsection{Strong polymer-substrate friction}

In this limit one has to take into account the specific geometry of an AFM
experiment, shown in Fig. \ref{Geometry} (right panel). For simplicity, an
infinite friction of the polymer at the surface is assumed. The adsorbed polymer
portion may be considered as a two-dimensional self-avoiding chain comprising
$N-n$ segments. The last contact point (marked as $c$ in Fig. \ref{Geometry})
can move due to adsorption or desorption elementary events. In  Ref. \cite{Netz}
this was classified as the {\it sticky} case. In Fig. \ref{Geometry}, $D$ is the
distance from the cantilever base to to the substrate, $R_{z}$ is the height of
the cantilever tip above the substrate, and $R$ is the distance between the
cantilever tip and the contact point $c$. Eventually, $R_{x}$ is the lateral
distance between cantilever base and the contact point $c$. One may assume that
initially the desorbed  portion of $n$ segments has occupied a distance of
$R_{x}$ which, due to self-avoiding $2D$-configurations of an adsorbed chain,
equals $R_{x}^2 \approx b^2 n^{2\nu}$ (where $\nu = 3/4$).

The specific geometry of the AFM experiment in  the case of strong
polymer-substrate friction (shown in Fig. \ref{Geometry}) brings about changes
only in the cantilever partition function, i.e., instead of Eq.
(\ref{Cantilever}), one has
\begin{eqnarray}
 \Xi_{\rm can} (D, R) &=& \exp \left[ - \dfrac{k_c}{2 k_BT} (D - R_{z})^2
\right].
\nonumber\\
&=& \exp \left[ - \dfrac{k_c}{2 k_BT} \left(D - \sqrt{R^2 - b^2
n^{2\nu}}\right)^2
\right]
\label{Cantilever_1}
\end{eqnarray}
As a result, the total partition function in this case is given by
\begin{eqnarray}
 \Xi_{\rm tot} (D) = \sum_{n=0}^{N} \: \exp\left[
 \epsilon (N - n)\right] \: \int\limits_{b n^{\nu}}^{b n} \: d R \; \exp \left[
n \: {\cal G} ({\widetilde f}) \right]
\: \exp \left[ - \dfrac{k_c}{2 k_BT} \left(D - \sqrt{R^2 - b^2 n^{2 \nu} }
\right)^2 \right] \theta (D - \sqrt{R^2 - b^2 n^{2 \nu}})
\label{Total_Partition_Function}
\end{eqnarray}
where again the variable  ${\widetilde f}$  should be excluded in favor of
$R/b n$  by means of the relation ${\widetilde f} = {\cal L}^{-1} (R/b n)$. In
Eq. (\ref{Total_Partition_Function}) the following constraints
\begin{eqnarray}
 b n^{\nu} &<& R < b n ,\nonumber\\
R_z &=& \sqrt{R^2 - b^2 n^{2 \nu} } < D ,
\end{eqnarray}
have been taken into account.

The corresponding free energy functional in terms of dynamical variables $n$
and  $R$ has the following form
\begin{eqnarray}
 {\cal F} (n, R) = - k_BT \epsilon (N - n) - k_BT n {\cal G} ({\widetilde f}) +
\dfrac{k_c}{2 } \left(D - \sqrt{R^2 - b^2 n^{2 \nu}} \right)^2 .
\label{Free_Energy}
\end{eqnarray}

The average force, which is measured in AFM-experiments, is given by
\begin{eqnarray}
 \langle f_{z} \rangle = \dfrac{k_c}{ \Xi_{\rm tot} (D) } \:  \sum_{n=0}^{N}
\: \: \int\limits_{b n^{\nu}}^{b n}
\: d R \;\left(D - \sqrt{R^2 - b^2 n^{2 \nu}}\right) \:\:
&&\exp\left[\epsilon (N - n) +  n \: {\cal G} ({\widetilde f})
  - \dfrac{k_c}{2 k_BT} \left(D - \sqrt{R^2 - b^2 n^{2 \nu}
}\right)^2 \right] \nonumber\\
& &\times \theta \left( D - \sqrt{R^2 - b^2 n^{2 \nu}}\right) .
\label{Average_Force_2}
\end{eqnarray}

\section{Dynamics of desorption}
\label{sec_dynamics}

 In our recent paper \cite{Paturej} we have studied a single polymer
force-induced desorption  kinetics
 by making use
of the notion of tensile blobs as well as by means of Monte
Carlo and Molecular Dynamics simulations.
It was clearly demonstrated that the total desorption time $< \tau_d>$
scales with polymer length $N$ as  $< \tau_d> \propto N^2$.

In order to treat a realistic AFM experiment in which the cantilever-substrate
distance changes with constant velocity $v_c$, i.e., $D(t) = D_0 + v_c t$,  one
has to consider the AFM tip dynamics. With this in mind, we will develop a
coarse-grained stochastic model based on the free-energy functional
Eq.~(\ref{Free_Energy_Frictionless}). Before proceeding any further, we
need to define the adsorption-desorption potential profile $F_{\rm ads}(n)$.
This plays the role of the  potential  of mean force (PMF) which depends  on
$n$.

\subsection{Stochastic Model}
\label{SM}

In the Helmholtz free-energy functional ${\cal F} (n, R)$, given by
Eq.~(\ref{Free_Energy_Frictionless}) and Eq.~(\ref{Free_Energy}), the free
energy of the adsorbed portion is given by a simple contact potential, $F_{\rm
ads} = - k_BT \epsilon (N - n)$, where $n$ is an integer number in the range $0
\le n \le N$. Considering desorption dynamics (see below), we will treat $n$ as
a continuous variable with a corresponding adsorption-desorption energy profile
satisfying the following conditions:
\begin{enumerate}
 \item For integer $n$-values the energy profile has
minima whereby we use the contact potential $ F_{\rm ads} (n) = - k_BT \epsilon
(N - n)$.
\item For half-integer values of $n$ the adsorption potential goes over
maxima.
\item The activation barrier for monomer desorption, $\Delta E^{+}=
F_{\rm ads} (n + 1/2) - F_{\rm ads} (n)$, and the corresponding adsorption
activation barrier, $\Delta E^{-} =  F_{\rm ads} (n + 1/2) - F_{\rm ads} (n +
1)$, are proportional to the adsorption strength $\epsilon$ of the substrate
whereby $\Delta E^{+} > \Delta E^{-}$.
\item The adsorption-desorption energy profile
satisfies the boundary conditions: $F_{\rm ads} (0) = - k_BT \epsilon N $ (a
fully adsorbed chain), and $F_{\rm ads} (N) = 0$ (an entirely detached chain).
\end{enumerate}

One may show that the following energy profile, given as
\begin{eqnarray}
 F_{\rm ads} (n) = T \epsilon \left\lbrace  1 + \cos[(2 n + 1) \pi]   + n
\right\rbrace - k_BT \epsilon N,
\label{Adsorption_Profile}
\end{eqnarray}
meets the conditions (1) - (4).

The  minima and maxima of Eq. (\ref{Adsorption_Profile}) are located in the
points defined by $\sin [(2 s + 1)\pi] = 1/2\pi$ with $s$ denoting the
{\em continuous} index of a monomer.

\begin{figure}[ht]
\begin{center}
 \includegraphics[width =12.0cm]{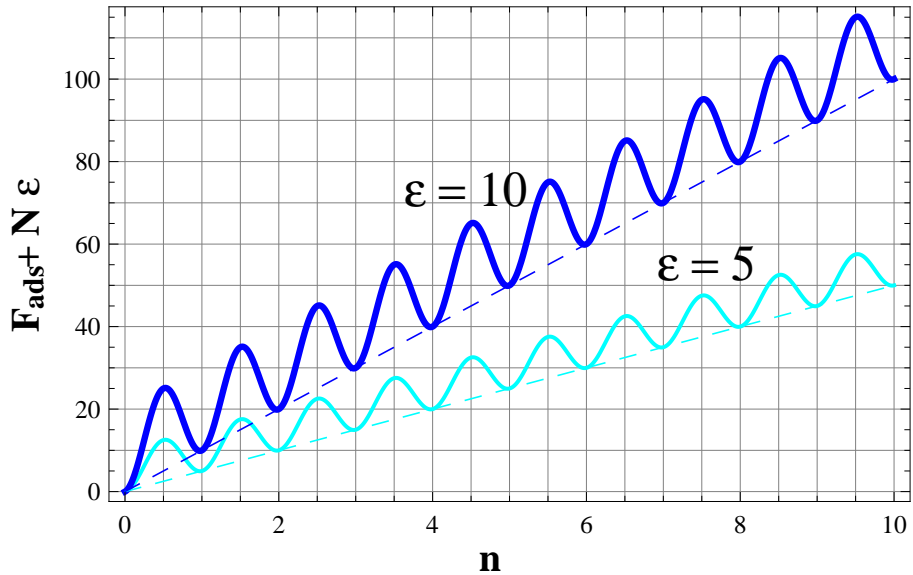}
\caption{Adsorption-desorption potential profiles for two different strengths
$\epsilon =
5$ and $\epsilon = 10$ of surface attraction. The integer $n$-values correspond
to minima whereas at half-integer $n$-values the potential has local maxima. The
dashed lines denote the corresponding contact potential $F_{\rm ads} = -
\varepsilon (N -n)$ which has been used in Sec. \ref{Second_Section}.}
\label{Pealing_Potential}
\end{center}
\end{figure}
As a result,
\begin{eqnarray}
 s  = \begin{cases}
                  \dfrac{1}{2\pi} \arcsin \left(\dfrac{1}{2 \pi}\right) + k -
\dfrac{1}{2} &\mbox{for}  \quad  k = 1, 2, \dots \\
\\
                 - \dfrac{1}{2\pi} \arcsin \left(\dfrac{1}{2 \pi}\right) + n
&\mbox{for} \quad n = 0, 1, 2, \dots
                \end{cases}
\label{Max_Min}
\end{eqnarray}

In Eq. (\ref{Max_Min}) the first term, $(1/2\pi) \arcsin (1/2\pi) \approx
0.025$, is very small and could be neglected. Thus, the minima and maxima
are located at the integer and half integer points respectively (see Fig.
\ref{Pealing_Potential})

In order to calculate the activation barriers, we determine first $F_{\rm ads}
(s)$ at the half-integer points, i.e.,
\begin{eqnarray}
 F_{\rm ads} (n + 1/2) =  k_BT \epsilon  (n + 5/2) - k_BT \epsilon N,
\end{eqnarray}
as well as at the integer points
\begin{eqnarray}
 F_{\rm ads} (n) &=&  k_BT \epsilon  n - k_BT \epsilon N \nonumber\\
F_{\rm ads} (n + 1) &=&  k_BT \epsilon  (n + 1)  - k_BT \epsilon N.
\end{eqnarray}

Therefore, the activation barriers for the detachment, $\Delta
E^{+}$, and adsorption, $\Delta E^{-}$, are given by
 \begin{eqnarray}
  \Delta E^{+} &=& F_{\rm ads} (n + 1/2) - F_{\rm ads} (n) =  \dfrac{5}{2} k_BT
\epsilon\nonumber\\
\Delta E^{-} &=& F_{\rm ads} (n + 1/2) - F_{\rm ads} (n + 1) = \dfrac{3}{2} k_BT
\epsilon
 \end{eqnarray}
i.e., $\Delta E^{+} > \Delta E^{-}$. Finally, one may readily see that $F_{\rm
ads}(0) = - k_BT \epsilon N$ and $F_{\rm ads} (N) = 0$ which is in line with
condition (iv).

The total Helmholtz free energy for the frictionless substrate model is given by
\begin{eqnarray}
 {\cal F} (n, R) = k_BT \epsilon \left\lbrace  1 + \cos[(2 n + 1) \pi]   + n
\right\rbrace - k_BT \epsilon N - k_BT n {\cal G} ({\widetilde f}) +
\dfrac{k_c}{2 } \left(D - R \right)^2 ,
\label{Free_Energy_Frictionless_Dynamics}
\end{eqnarray}
whereas for the strong polymer-substrate friction model we have
 \begin{eqnarray}
{\cal F} (n, R) = k_BT \epsilon \left\lbrace  1 + \cos[(2 n + 1) \pi]   + n
\right\rbrace - k_BT \epsilon N - k_BT n {\cal G} ({\widetilde f}) +
\dfrac{k_c}{2 } \left(D - \sqrt{R^2 - b^2 n^{2 \nu}} \right)^2 .
\label{Free_Energy_Dynamics}
\end{eqnarray}

These Helmholtz free energy functions govern the dissipative process which is
described  by  the stochastic (Langevin) differential equations
\begin{eqnarray}
 \dfrac{\partial n}{\partial t} &=& - \lambda_{n} \: \dfrac{\partial}{\partial
n} \:
{\cal F} (n, R) + \xi_{n} (t) \nonumber\\
\dfrac{\partial R}{\partial t} &=&  - \lambda_{R} \: \dfrac{\partial}{\partial
R}
 {\cal F} (n, R) + \xi_{R} (t)
\label{Onsager}
\end{eqnarray}
where $\lambda_{n}$ and $\lambda_{R}$ are the Onsager coefficients. The random
forces $\xi_{n} (t)$ and $\xi_{R} (t)$ describe Gaussian noise with
means and correlators given by
\begin{eqnarray}
 \langle \xi_{n} (t) \rangle &=& \langle \xi_{R} (t) \rangle = 0  \nonumber\\
\langle \xi_{n} (t)  \xi_{n} (0) \rangle &=& 2 \lambda_{n} k_BT \delta (t)
\nonumber\\
\langle \xi_{R} (t)  \xi_{R} (0) \rangle &=& 2 \lambda_{R} k_BT \delta (t)
\label{Random_Forces}
\end{eqnarray}
Equations (\ref{Onsager}) are usually referred to as the Onsager equations
\cite{Puri}.

The set of stochastic differential equations, Eq.(\ref{Onsager}) can be treated
by a time integration scheme. Each realization $(l)$ of the solution provides a
time evolution of $n^{(l)} (t)$ and $R^{(l)} (t)$. In order to get mean values
of the observables, these trajectories should be averaged over many independent
runs $l =1, 2, \dots  {\cal N}$. For example, in order to obtain the average
force, Eq. (\ref{Average_Force_2}), one should average over the runs
\begin{eqnarray}
 \langle f_{z} (t) \rangle = \dfrac{k_c}{\cal N} \: \sum_{l = 1}^{\cal N} \:
\left(D(t)  - \sqrt{[R^{(l)}(t)]^2 - b^2 [n^{(l)}(t)]^{2 \nu}}\right)
\end{eqnarray}

\subsection{Thermodynamic forces}

The thermodynamic forces which arise in Eq. (\ref{Onsager}), i.e.,
$f_{n} \stackrel{\rm def}{=} - \frac{\partial {\cal F} (n, R}{\partial n}$ and
$f_{R} \stackrel{\rm def}{=} - \frac{\partial {\cal F} (n, R)}{\partial R}$,
could be calculated explicitly.
For example, for the free energy function, Eq.
(\ref{Free_Energy_Frictionless_Dynamics}), one  has
\begin{eqnarray}
 f_{n} &=& - k_BT \epsilon \left\lbrace   1 - 2 \pi \sin [(2 n + 1) \pi ]
\right\rbrace  + k_BT \: {\cal G} ({\widetilde f}) + k_BT \: n  \: {\cal
G}^{\prime} ({\widetilde f}) \: \left( \dfrac{\partial {\widetilde f}}{
\partial n}\right)_{R} \nonumber\\
&=& - k_BT \epsilon  \left\lbrace   1 - 2 \pi \sin [(2 n + 1) \pi ]
\right\rbrace    + k_BT \: {\cal G} ({\widetilde f}) - \dfrac{k_BT \: R}{b\: n}
\:
\dfrac{{\cal G}^{\prime} ({\widetilde f})}{{\cal L}^{\prime} ({\widetilde f})},
\label{f_n_1}
\end{eqnarray}
where we have used  (recall that  $R/b n = {\cal L} ({\widetilde f})$)
\begin{eqnarray}
 \left( \dfrac{\partial {\widetilde f}}{\partial n}\right)_{R}  = -
\dfrac{R /b n^2}{ {\cal L}^{\prime} ({\widetilde f})}. \nonumber
\end{eqnarray}

On the other hand, a direct calculation shows that
\begin{eqnarray}
 {\cal G}^{\prime} (x) &=& \dfrac{x}{[\sinh (x)]^2} \: - \:
\dfrac{1}{x}\nonumber\\
{\cal L}^{\prime} (x)  &=& \dfrac{1}{x^2} \: - \: \dfrac{1}{[\sinh (x)]^2}
\end{eqnarray}
so that
\begin{eqnarray}
 \dfrac{{\cal G}^{\prime} (x)}{{\cal L}^{\prime} (x)} = - x.
\label{Relationship}
\end{eqnarray}
Thus, for the force $f_n$, given by Eq. (\ref{f_n_1}), one has
\begin{eqnarray}
 f_{n} = - k_BT  \epsilon \left\lbrace  1 - 2 \pi \sin [(2  n + 1 ) \pi]
\right\rbrace + k_BT \dfrac{R \: {\widetilde f}}{b\: n} + k_BT
\: {\cal G} ({\widetilde f}).
\label{Force_Frictionless}
\end{eqnarray}

In the strong friction case, Eq. (\ref{Free_Energy_Dynamics}) leads to a more
complicated expression for the thermodynamic force:
\begin{eqnarray}
 f_{n} = - k_BT  \epsilon \left\lbrace   1 - 2 \pi \sin [(2  n + 1 ) \pi]
\right\rbrace  + k_BT \dfrac{R \: {\widetilde f}}{b\: n} + k_BT
\: {\cal G} ({\widetilde f})   + \: \nu \; b^2 \: k_c \; \left(
\dfrac{D}{\sqrt{R^2 - b^2 n^{2 \nu}}} - 1 \right).
\label{Force_Strong_Friction}
\end{eqnarray}

For  $f_{R} \stackrel{\rm def}{=} -\partial {\cal F} (n, R)/ \partial
R$ one obtains
\begin{eqnarray}
 f_{R} = k_BT \: n \: {\cal G}^{\prime} ({\widetilde f}) \: \left(
\dfrac{\partial {\widetilde f}}{\partial R}\right)_{n}  + k_c (D - R)
= \dfrac{k_BT}{b} \: \dfrac{{\cal G}^{\prime} ({\widetilde f})}{{\cal
L}^{\prime} ({\widetilde f})} \: + \:  k_c (D - R)
\end{eqnarray}
where we have used
\begin{eqnarray}
 \left( \dfrac{\partial {\widetilde f}}{\partial R}\right)_{n}  =
\dfrac{1 /b n}{ {\cal L}^{\prime} ({\widetilde f})}. \nonumber
\end{eqnarray}
Taking into account Eq. (\ref{Relationship}), one finally derives
\begin{eqnarray}
 f_{R} = -  \dfrac{k_BT \: {\widetilde f}}{b} \: + \:  k_c (D - R).
\label{f_R_1}
\end{eqnarray}

For the model, given by Eq. (\ref{Free_Energy_Dynamics}), the corresponding
force reads
\begin{eqnarray}
 f_{R} = -  \dfrac{k_BT \: {\widetilde f}}{b} \: + \:  k_c \; R \:
\left(\dfrac{D}{\sqrt{R^2 - b^2 n^{2 \nu}}} -  1 \right).
\label{f_R_2}
\end{eqnarray}

Finally,  the variable ${\widetilde f}$ should be expressed in terms of $R/b n$
by making use of the relationship ${\widetilde f} = {\cal L}^{-1} (R/b n)$,
where ${\cal L}^{-1} (x)$ is the inverse Langevin function. A very good
approximation for the inverse Langevin function, published in Ref.
\cite{Cohen}, is given by
 \begin{eqnarray}
  {\widetilde f} = {\cal L}^{-1} (R/b n)
\approx& \dfrac{R}{b n} \: \left[ \dfrac{3 - \left(\dfrac{R}{b n}\right)^2}{1 -
\left(\dfrac{R}{b n}\right)^2} \right]
 \label{Interpolation}
\end{eqnarray}

\subsection{Quasistationary approximation}
\label{Quasistationary_approx}

It could be shown that for a strongly stretched desorbed portion of the polymer
chain, the $R$ variable rapidly relaxes to its {\bf quasi-stationary} value (see
Appendix \ref{App}). In other words, $R$ can quickly adjust to the slow
evolution of $n$ (governed by the Kramers process).  In this quasi-stationary
approximation $f_R = 0$, and from Eq. (\ref{f_R_1}) one has $k_c (D - R) = k_BT
{\widetilde f}/b$, so that the following nonlinear equation for $R$ emerges
\begin{eqnarray}
 k_c [D(t)  - R] = \dfrac{k_BT}{b} \: {\cal L}^{-1} \left(\frac{R}{nb}\right)
\label{L^{-1}}
\end{eqnarray}
This could be represented as
\begin{eqnarray}
 G(n, R; t) \stackrel{\rm def}{=} \dfrac{R}{b n} - {\cal L} \left( \dfrac{
k_c b [D (t) - R]}{k_BT} \right) = 0,
\label{G}
\end{eqnarray}
i.e., the height $R$ is instantaneously coupled to the number of desorbed
beads, $n$. Inserting Eq. (\ref{Interpolation}) into Eq. (\ref{L^{-1}}), one
obtains
\begin{eqnarray}
 P(n, R; t) \stackrel{\rm def}{=} \dfrac{k_c b [D(t) - R]}{k_BT} - \left(
\dfrac{R}{b n}\right) \: \left[ \dfrac{3 - \left(\dfrac{R}{b n}\right)^2}{1 -
\left(\dfrac{R}{b n}\right)^2} \right]  = 0 .
\label{P}
\end{eqnarray}

The Onsager equation for the slow variable $n$ is given as
\begin{eqnarray}
 \dfrac{\partial n}{\partial t} &=& \lambda_n \: f_n (n, R) + \xi_n
(t)\nonumber\\
&=& \lambda_n \: [- k_BT  \epsilon \left\lbrace  1 - 2 \pi \sin [(2  n + 1 )
\pi]
\right\rbrace + k_BT \dfrac{R \: {\widetilde f}}{b\: n} + k_BT
\: {\cal G} ({\widetilde f})]  + \xi_n (t)
\label{Kramers_Process}
\end{eqnarray}
where ${\widetilde f}$ is determined by Eq. (\ref{Interpolation}).

Eventually, we get a system of so-called semi-explicit
differential-algebraic equations (DAE) \cite{Ascher}
\begin{eqnarray}
 \dfrac{\partial n}{\partial t} &=& \lambda_n \: f_n (n, R) + \xi_n
(t)\nonumber\\
 0 &=&  G(n, R; t)
\label{DAE}
\end{eqnarray}
In this particular form of DAE one can distinguish between the differential
variable $n(t)$ and the algebraic variable $R(t)$. Eq. (\ref{DAE}) can be
solved numerically by making use of an appropriate Runge - Kutta (RK)
algorithm, as shown in the Appendix \ref{R-K}.

\subsection{Results}
\label{Res}

\begin{figure}[ht]
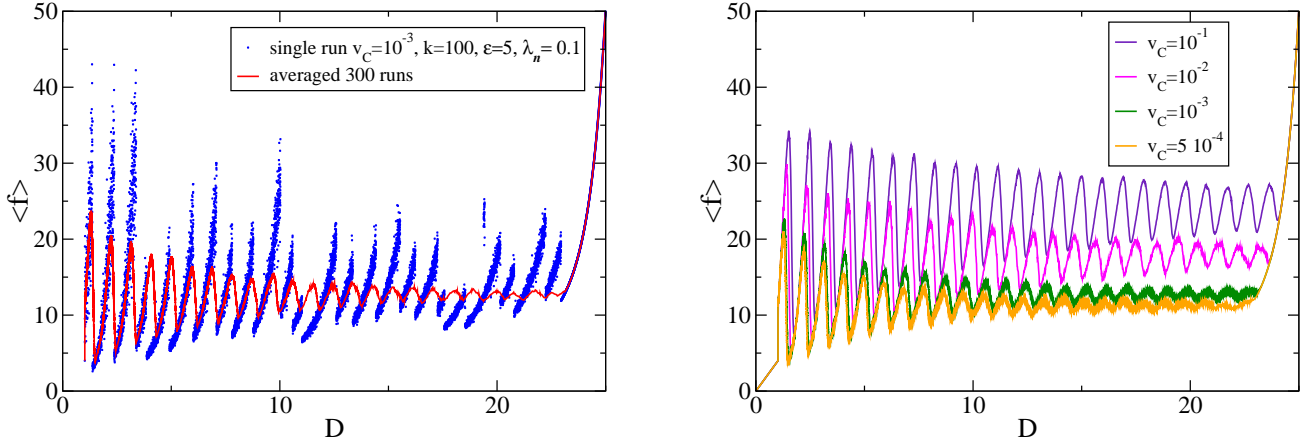

\begin{center}
\vspace{0.7cm}
\includegraphics[width =8cm]{average_vs_single.eps}
\hspace{1.0cm}
\includegraphics[width =8cm]{force_D_e5.eps}

\caption{(left panel) The effect of averaging over many desorption events. The
blue line shows a single run, whereas the red line demonstrates the result of
averaging over $300$ runs. The detachment velocity $v_c = 100$, the cantilever
spring $k_c = 100$, adsorption energy $\epsilon = 5$, and the Onsager
coefficient $\lambda = 0.1$. (right panel) The dynamic force-displacement
diagram for $\epsilon = 5$ and different detachment velocities $v_c$ (see
legend). }
\label{Averaging}
\end{center}
\end{figure}

We have solved numerically our stochastic model, given by Eq.
(\ref{Kramers_Process}) and Eq. (\ref{G}), for the case of frictionless
substrate. To this end we used the  second order  Runge - Kutta (RK) algorithm
for
stochastic differential-algebraic equations (see Appendix \ref{R-K} for more
details). The advantage of the stochastic differential equations approach as
compared to the Master Equation method \cite{Kreuzer_3} is that the former one
gives a more detailed (not averaged) dynamic information corresponding to each
individual force-displacement trajectory (as is often in an experiment). The
result of averaging over $300$ runs is shown in Fig. \ref{Averaging} (left).

Fig.~\ref{Averaging} (right panel)  shows the resulting force - displacement
diagram for $\epsilon = 5$ and different detachment velocities. It it worth
noting that the ''sawtooth'' pattern can be seen for all investigated detachment
velocities ranging between $v_c = 5 \times 10^{-4}$ and $v_c = 10^{-2}$. For
larger velocities the plateau height of the force grows substantially.
In other words, the mean detachment force increases as
the AFM-tip velocity gets higher and the bonds stretching between successive
monomers becomes stronger.

\begin{figure}[ht]
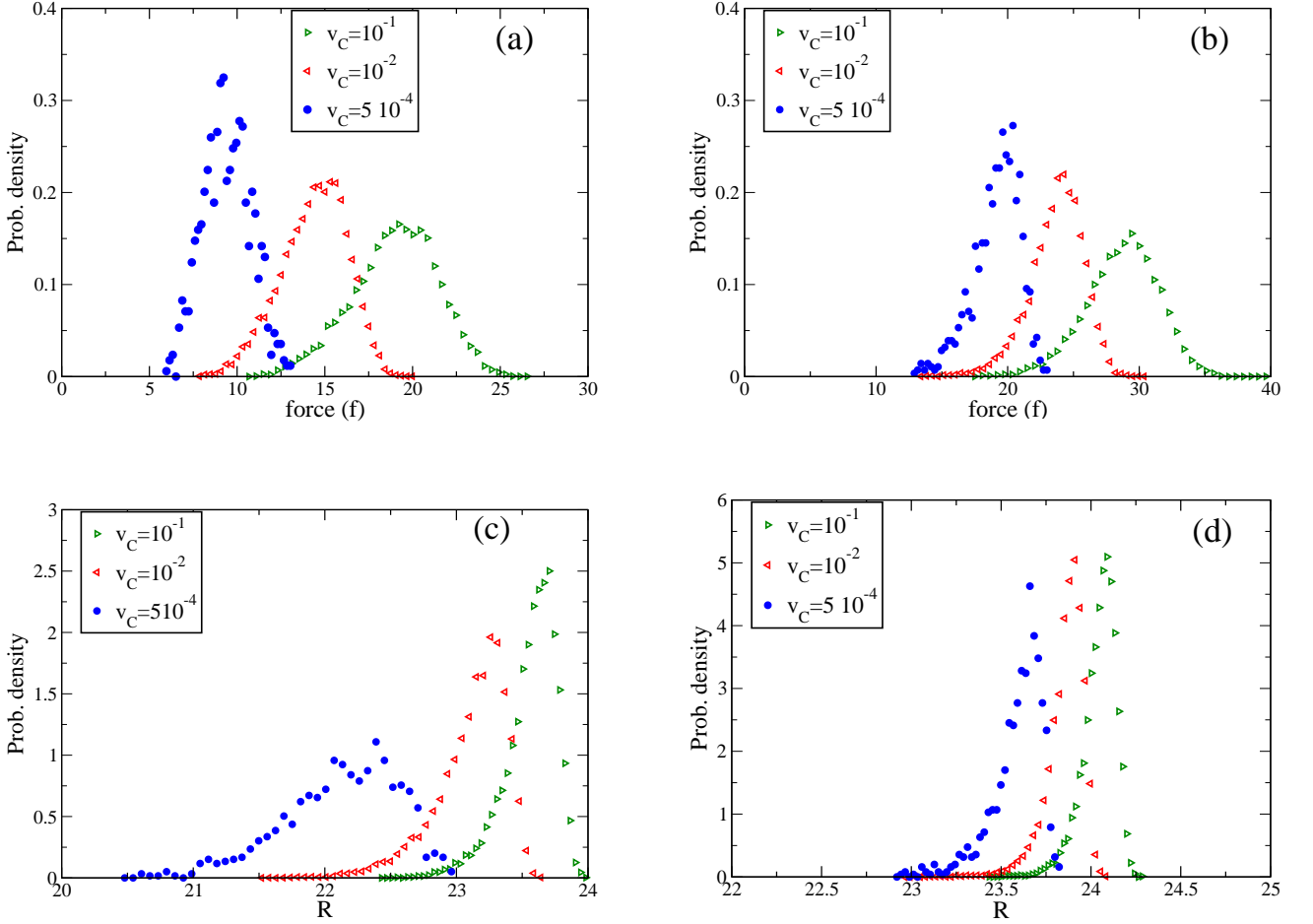

\begin{center}
\vspace{0.5cm}
\includegraphics[width =8.0 cm]{fprobforce_ep5.eps}
\hspace{1cm}
\includegraphics[width =8.0 cm]{fprobforce_ep8.eps}

\vspace{0.9cm}

\includegraphics[width =8.0 cm]{RprobfR_ep5.eps}
\hspace{1cm}
\includegraphics[width =8.0 cm]{RprobfR_ep8.eps}

\caption{The normalized PDF for detachment forces at $\epsilon = 5$ (a), and
$\epsilon = 8$ (b), as well as for different detachment velocities $v_c$ (shown
in the legends). The PDFs for the detachment distance $R$ of cantilever tip at
$\epsilon = 5$ (c) and $\epsilon = 8$ (d).}
\label{PDF_Force}
\end{center}
\end{figure}

We have also studied the detachment force behavior as well as that of the
cantilever tip distance from the substrate at the moment of a full detachment,
(i.e. when $n = N$), by repeating the detachment procedure $10^4$ times and
plotting the probability distribution functions (PDF) for different  adsorption
energies  $\epsilon$ and detachment velocities $v_c$ - Fig. \ref{PDF_Force}. As
one can see from Fig. \ref{PDF_Force}a,~b,  both the average and the dispersion
of detachment force grow with $v_c$ which agrees with findings for reversible
(i.e., when a broken bond can rebind) bond-breaking dynamics \cite{Diezemann}.
In contrast, the mean cantilever tip
distance $R$ variance decreases and its average value increases with growing
$v_c$ (cf. Fig. \ref{PDF_Force}c,~d).

\begin{figure}[ht]
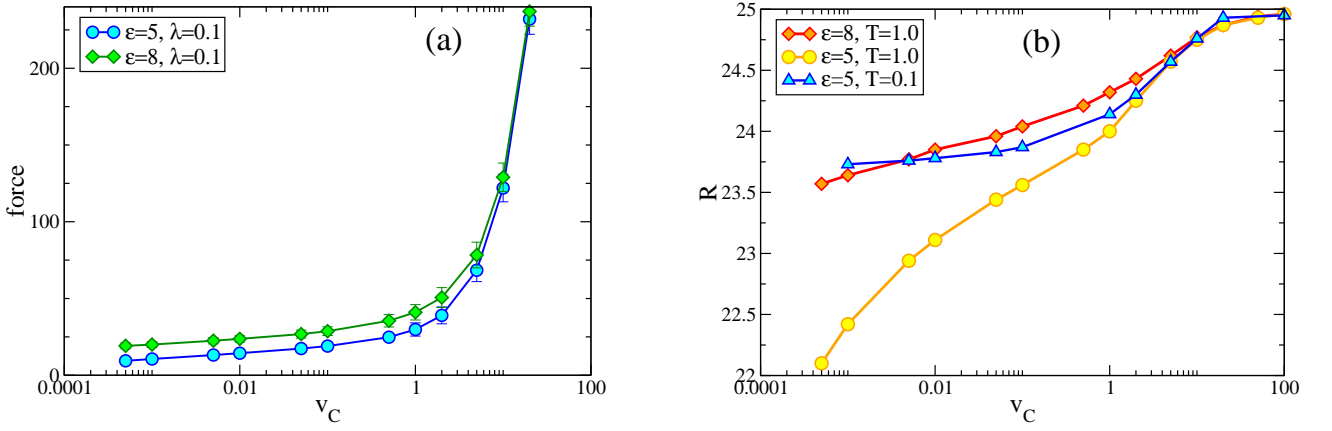

\begin{center}
\vspace{0.7cm}
\includegraphics[width =8.0 cm]{force_vc_pap_REV.eps}
\hspace{1cm}
\includegraphics[width =8.0 cm]{R_vc_pap_REV.eps}

\caption{The average detachment force (a), and cantilever tip distance (b),
vs detachment velocity $v_c$ for two different adsorption energies $\epsilon =
5$ and $\epsilon = 8$. In addition, the distance $R$ is plotted in (b) for two
different temperatures, $T=1.0$ and $T=0.1$, both at the same adsorption
strength $\epsilon=5.0$. Evidently, with growing $T$ the elasticity of the coil
decreases, the coil itself becomes stiffer and therefore detaches from the
substrate at lwoer height $R$. This entropic effect is well expressed at
sufficiently low pulling velocity only.}
\label{Force_Average}
\end{center}
\end{figure}

The average detachment force dependence on cantilever velocity $v_c$ is a
widely covered subject in the literature in the context of biopolymers unfolding
\cite{Kreuzer_4,Eom_1,Eom_2} or forced separation of two adhesive surfaces
\cite{Leckband,Seifert,Diezemann}. Figure~\ref{Force_Average}a, which shows the
result of our calculations, has the characteristic features discussed also in
ref. \cite{Leckband}.  One observes a well expressed crossover from a
shallow-slope for relatively small detachment rates to a steep-slope region as
detachment speed increases. One remarkable feature is that this crossover
practically does not depend on the adsorption energy $\epsilon$: the curve is
merely shifted upwards upon increasing of $\epsilon$. Therefore, the crossover
is not related to a competition between the Kramers rate and the cantilever
velocity but rather accounts for the highly nonlinear chain stretching as the
velocity $v_c$ increases. The corresponding detachment distance of the
cantilever tip $R$ (detachment height), Fig.~\ref{Force_Average}b, reveals a
specific sigmoidal shape in agreement with the results based on the Master
Equation \cite{Kreuzer_3}. At low velocities of pulling, $v_c$, when the chain
still largely succeeds in relaxing back to equilibrium during detachment, an
interesting entropy effect is manifested in Fig.~\ref{Force_Average}b: the
(effectively) stiffer coil at $T=1.0$ leaves the substrate at lower values of
$R$ than in the case of the colder system, $T=0.1$. As the pulling velocity
grows, however, this entropic effect vanishes and the departure from the
substrate is largely governed by the stretching of the bonds rather than of the
coil itself whereby the difference in behavior between $T=1.0$ and $T=0.1$
disappears.

\begin{figure}[ht]
\begin{center}
\vspace{0.7cm}
\includegraphics[width =8.5 cm]{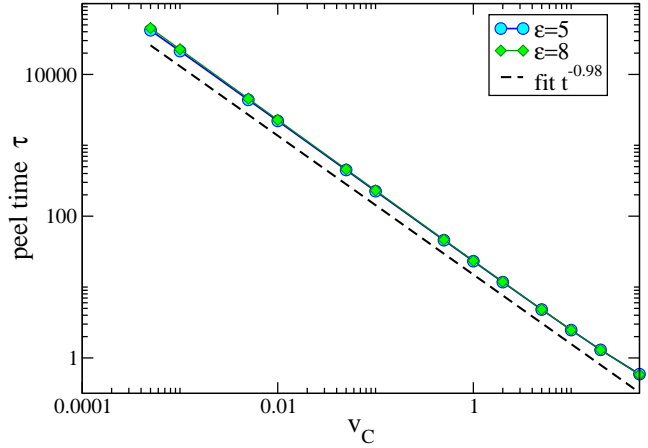}
\caption{The average  detachment (peeling) time $\tau_{\rm det}$ versus
detachment velocity $v_c$ for two different adsorption energies $\epsilon =
10$ and $\epsilon = 16$.  The inversely proportional dependence, $\tau_{\rm det}
\sim 1/v_c$, agrees well with previous theoretical findings
\cite{Seifert}.}
\label{Detachment_Time}
\end{center}
\end{figure}

Eventually, as it can be seen from Fig.~\ref{Detachment_Time}, the total
detachment (peel) time $\tau_{\rm det}$ vs. velocity $v_c$ relationship has a
well-defined power-law behavior, $\tau_{\rm det} \sim 1/v_c^{\alpha}$, with the
power $\alpha \approx 1$, in line with previous theoretical findings
\cite{Seifert}.

\section{MD simulations}
\label{sec_MD}
\subsection{The model}

In our MD-simulations we use  a coarse-grained model of a polymer chain of $N$
beads connected by finitely extendable elastic bonds. The bonded interactions in
the chain is described by the frequently used Kremer-Grest potential,
$V^{\mbox{\tiny KG}}(r)=V^{\mbox{\tiny FENE}}(r) + V^{\mbox{\tiny WCA}}(r)$.
The FENE (finitely extensible nonlinear elastic) potential is given by
\begin{equation}
V^{\mbox{\tiny FENE}}= -\frac 12 k r_0^2 \ln{\left[ 1 - \left(\frac
r{r_0}\right)^2 \right]}
\label{fene}
\end{equation}
with $k=30\,\epsilon/\sigma^2$ and $r_0=1.5\sigma$.

In order to allow properly for excluded volume interactions between bonded
monomers, the repulsion term is taken as Weeks-Chandler-Anderson (WCA) potential
(i.e., the shifted and truncated repulsive branch of the Lennard-Jones
potential,) given by
\begin{equation}
 V^{\mbox{\tiny WCA}}(r) = 4\epsilon\left[
(\sigma/ r)^{12} - (\sigma /r)^6 + 1/4
\right]\theta(2^{1/6}\sigma-r)
\label{wca}
\end{equation}
with $\theta(x)=0$ or $1$ for $x<0$, or $x\geq 0$, and $\epsilon=1$, $\sigma=1$.
The overall potential $V^{\mbox{\tiny KG}}(r)$ has a minimum at bond length
$r_{\mbox{\tiny bond}}\approx 0.96$. The nonbonded interaction between monomers
are taken into account by means of the WCA potential, Eq.~\ref{wca}. Thus, the
interactions in our model correspond to good solvent conditions.

The substrate in the present investigation is considered simply as a
structureless adsorbing plane, with a Lennard-Jones potential acting with
strength $\epsilon_s$ in the perpendicular $z$--direction, $V^{\mbox{\tiny
LJ}}(z)=4\epsilon_s[(\sigma/z)^{12} - (\sigma/z)^6]$. In our simulations we
consider as a rule the case of strong adsorption $\epsilon_s/k_BT=5\div20$,
where $k_BT$ is a temperature of Langevin thermal bath described below.


The dynamics of the chain is obtain by solving the Langevin equations of motion
for the position $\mathbf r_n=[x_n,y_n,z_n]$ of each bead in the chain,
\begin{equation}
m\ddot{\mathbf r}_n = \mathbf F_n^{j} + \mathbf F_n^{\mbox{\tiny WCA}}
-\gamma\dot{\mathbf r}_n + \mathbf R_n(t) \qquad
(1,\ldots,N)
\label{langevin}
\end{equation}
which describes the Brownian motion of a set of bonded particles.

The influence of solvent is split into slowly evolving viscous force and rapidly
fluctuating stochastic force. The random Gaussian force $\mathbf R_n$ is related
to friction coefficient $\gamma=0.25$ by the fluctuation-dissipation theorem.
The integration step is $\tau=0.005$ and time in measured in units of
$\sqrt{m\sigma^2/\epsilon}$, where $m$ denotes the mass of the polymer beads,
$m=1$. In all our simulations the velocity-Verlet algorithm was used to
integrate equations of motion (\ref{langevin}).
\begin{figure}[ht]
\begin{center}
\vspace{0.5cm}
\includegraphics[width=8 cm]{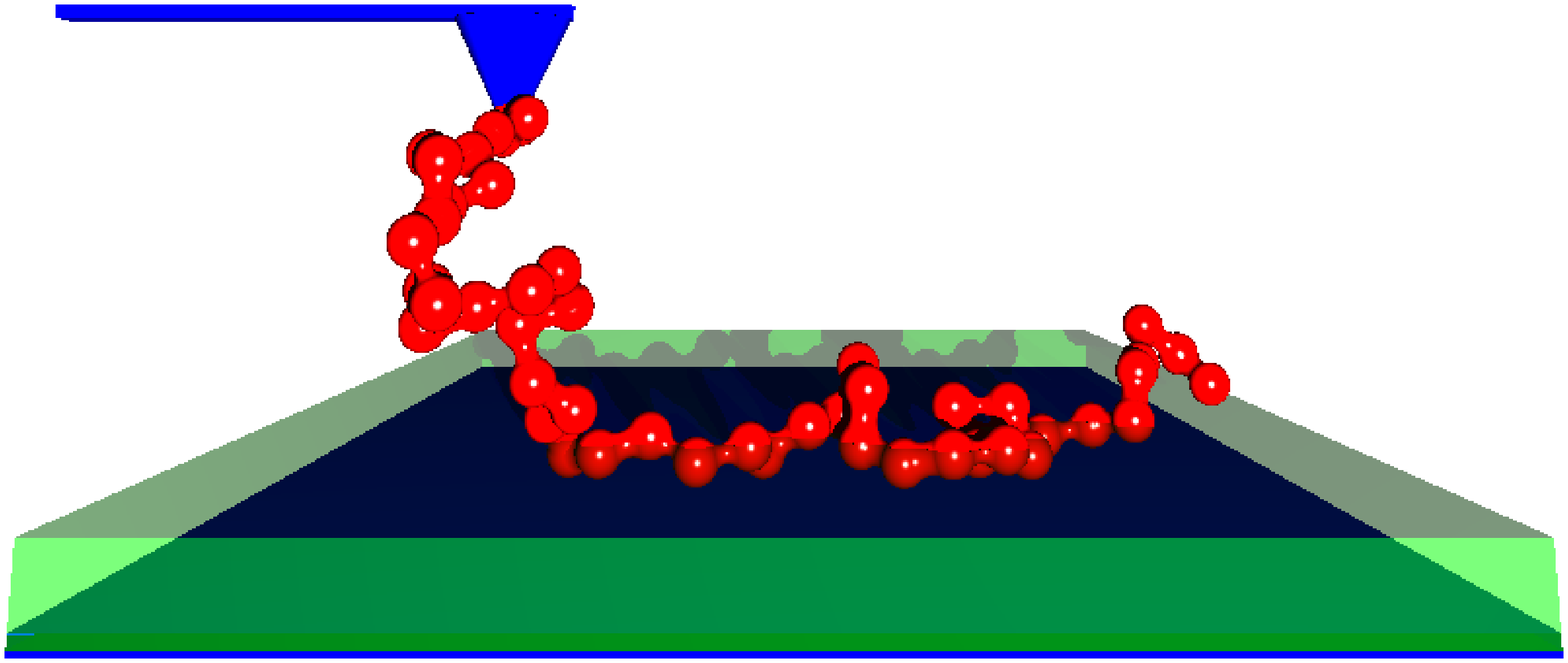}
\hspace{0.7cm}
\includegraphics[width=8 cm]{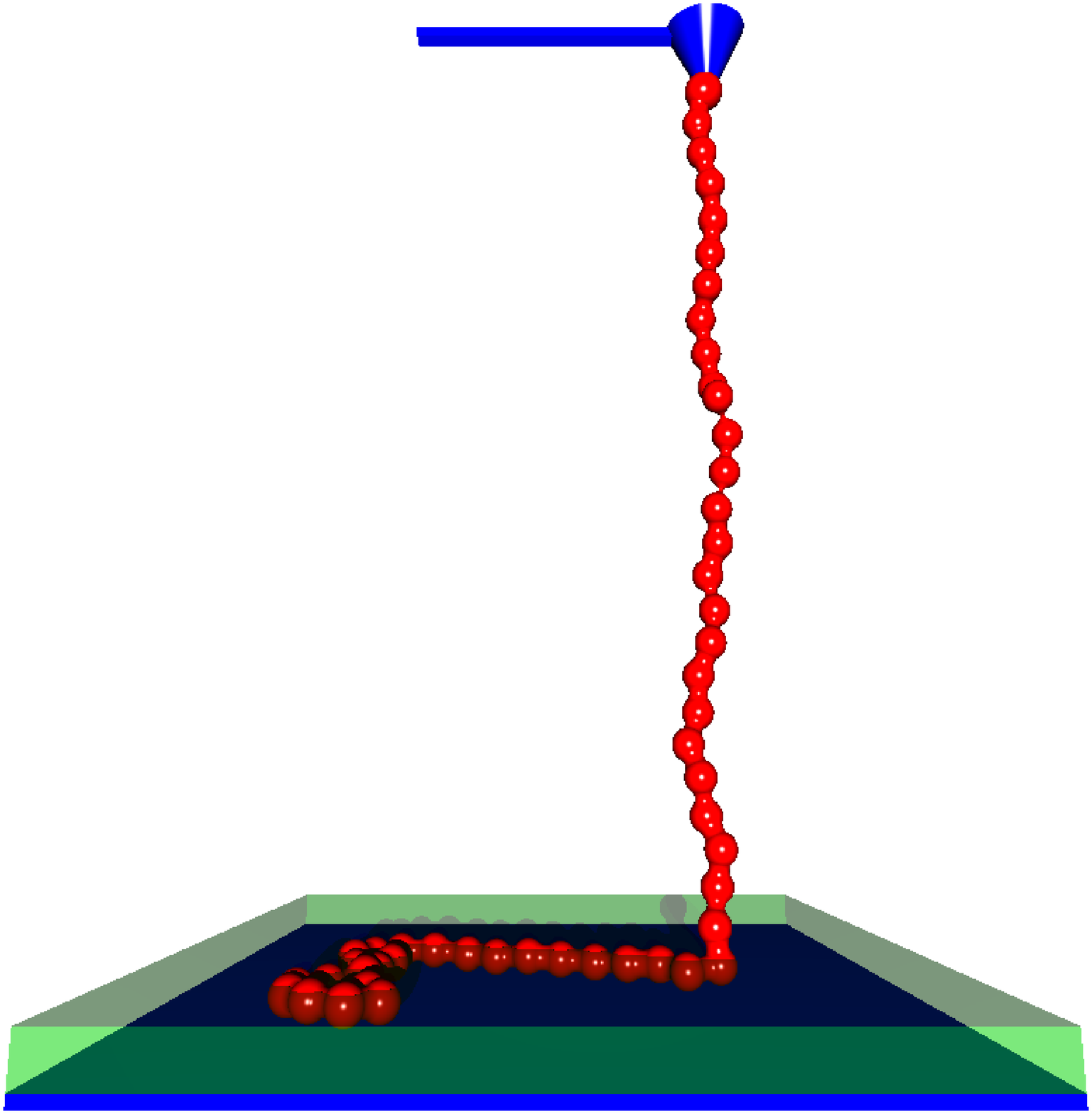}
\caption{Snapshots of a polymer chain with $N=50$ monomers during detachment
from a substrate at $\varepsilon = 2.5$ (left) and $\varepsilon = 20$
(right). Here $v_c = 0.0001$. The interaction range of the adsorption potential
is shaded (transparent) green. The cantilever tip is shown schematically in
blue. One may clearly see that the polymer chain is more relaxed (less
stretched) at $\varepsilon = 2.5$, and the adsorbed monomers do not stick
tightly to the surface but partially exit the range of surface adsorption. }
\label{fig_snapshots}
\end{center}
\end{figure}

The molecule is pulled by a cantilever at constant velocity $\mathbf V
=[0,0,v_c]$. The cantilever  is imitated by  two beads connected by harmonic
spring and attached to one of the ends of the chain. \footnote{This setup is
different than the one used by S. Iliafar {\it et al.} \cite{Jagota_3}); in
their study a harmonic spring was connected to a "big" monomer (with large
friction coefficient) on one side, and to a mobile wall on the other side. In
our case the harmonic spring  spans {\em two} beads.}

The mass of beads $m_c$, forming the cantilever, was set either to $m_c=1$ or to
$25$. The equilibrium size of this harmonic spring  was set to $0$ and the
spring constant was varied in the range $k_c=50 \div 400\, \epsilon/\sigma^2$.
The hydrodynamics radius $a$ of beads composing the cantilever  was varied by
changing the friction coefficient $\gamma_c = 0.25 \div 25$, taking into account
the Stokes' law, $\gamma_c = 6\pi\eta a$, where $\eta$ is the solvent viscosity.

Taking the value of the thermal energy $k_BT \approx 4.11\times 10^{-21}$~J at
$k_BT=300$~K, the typical Kuhn length of $\sigma = 1$~nm and the mass of
coarse-grained monomer as $m\approx 10^{-25}$~kg setups the unit of time in our
simulations which is given in $~10^{-12}\,{\text{s}}=1$~ps. The velocities
used in simulations are in units of $10^{-4}\div10^{-1}\,
\text{nm}/\text{ps}\approx 10^{-1}\div 10^2\, \text{m}/\text{s}$. Spring
constants of our cantilever  in real units are: $k_c=50\div400\,
k_BT/(\text{nm})^2=0.2\div1.6\, \text{N}/\text{m}$.

Two typical snapshots of a polymer chain during slow detachment from an
adsorbing substrate with different strengths of adsorption, $\varepsilon=2.5$
and $\varepsilon=20$ are shown in Fig.~\ref{fig_snapshots}. Evidently, the
chain is much more stretched for the strongly-attractive substrate where all
adsorbed monomers stick firmly to the surface.

\begin{figure}[ht]
\begin{center}
\vspace{0.5cm}
\includegraphics[scale=0.3]{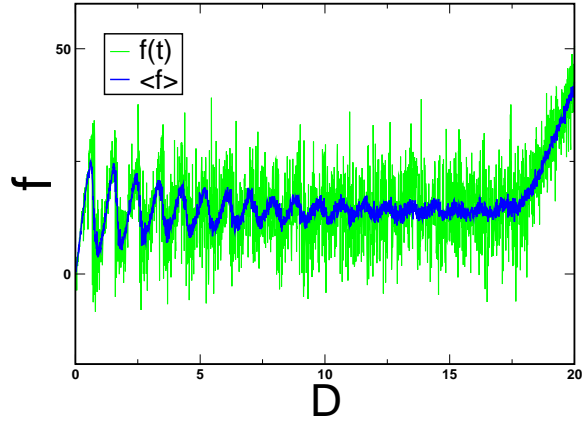}
\caption{ Comparison of averaged force measured at the cantilever vs single
realization of desorption experiment. Force $f$ at the cantilever was calculated
by monitoring extensions $\Delta z_c$ of a harmonic spring, i.e. $f=k_c\Delta
z_c$. Here $N=20$, $k_c=50\,\epsilon/\sigma^2$, $v_c=10^{-4}\,\sigma/\tau$ and
$\epsilon_s/k_BT=20$. $m_c=1$ and $\gamma_c=0.25$. }
\label{Average_1}
\end{center}
\end{figure}

\subsection{MD-results}
\label{MM}

As we have already seen in Sec. \ref{Res}, the averaging of the force profile
over many runs reveals the inherent sawtooth-structure of the force vs distance
dependence (see Fig.~\ref{Averaging}) which is otherwise overshaded by thermal
noise. Our MD-simulation result, depicted in Fig. \ref{Average_1}, show the
same
tendency against the noisy background of a single detachment event. Therefore,
for better clarity and physical insight, all our graphic results that are given
below result from such averaging procedure.

\begin{figure}[ht]
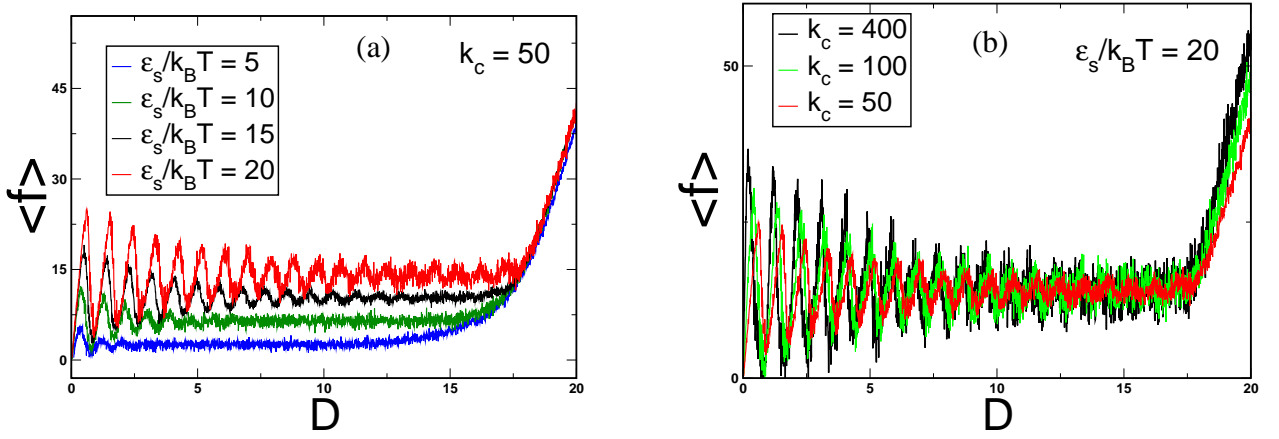

\begin{center}
\includegraphics[scale=0.3]{fVSD_eps.eps}
\hspace{1cm}
\includegraphics[scale=0.3]{fVSD_kspring.eps}
\caption{Time-averaged force $\langle f \rangle$ at the cantilever as a function
of a distance from the substrate $D$ during chain detachment. Here $N=20$,
$v_c=10^{-4}\,\sigma/\tau$, $m_c=1$ and $\gamma_c=0.25$. Presented results are
for different adsorption strengths $\epsilon$, as indicated in the legend, and
$k_c=50\epsilon/\sigma^2$ (a). Figure (b) presents results for $\epsilon = 20$
and different spring constants $k_c$. Each curve originates from $10^3$
independent simulations.}
\label{Force_vs_Distance_MD}
\end{center}
\end{figure}

Figure \ref{Force_vs_Distance_MD}a shows how adsorption energy $\epsilon$
affects the force $f$ vs distance $D$ relationship. Apparently, with
increasing $\epsilon$ the mean force (plateau height) is found to grow in
agreement with our
equilibrium theory results, given in Fig. \ref{Equilibrium_1}. As suggested by
our recent theory \cite{Bhattacharya_1,Bhattacharya_3}, the plateau height goes
up as $f_p \propto \epsilon^{1/2}$, or as $f_p \propto \epsilon$, for relatively
small or large $\epsilon$ values, respectively. The amplitude of spikes
increases with growing $\epsilon$ too, in line with the equilibrium findings
(see Fig.~\ref{Equilibrium_1}). Moreover, as found by  Jagota {\it et al.}
\cite{Jagota_1}, the amplitude of spikes follows an exponential law, $f_{\rm
amp} \propto \exp (\epsilon/n)$, where $n$ is the number of desorbed polymer
segments. On the other hand, the comparison of Fig.~\ref{Force_vs_Distance_MD}b
and Fig.\ref{Equilibrium} suggests that the stiffness of the cantilever spring
constant $k_c$ affects mainly the spike amplitude especially at large
$\epsilon$.

\begin{figure}[ht]
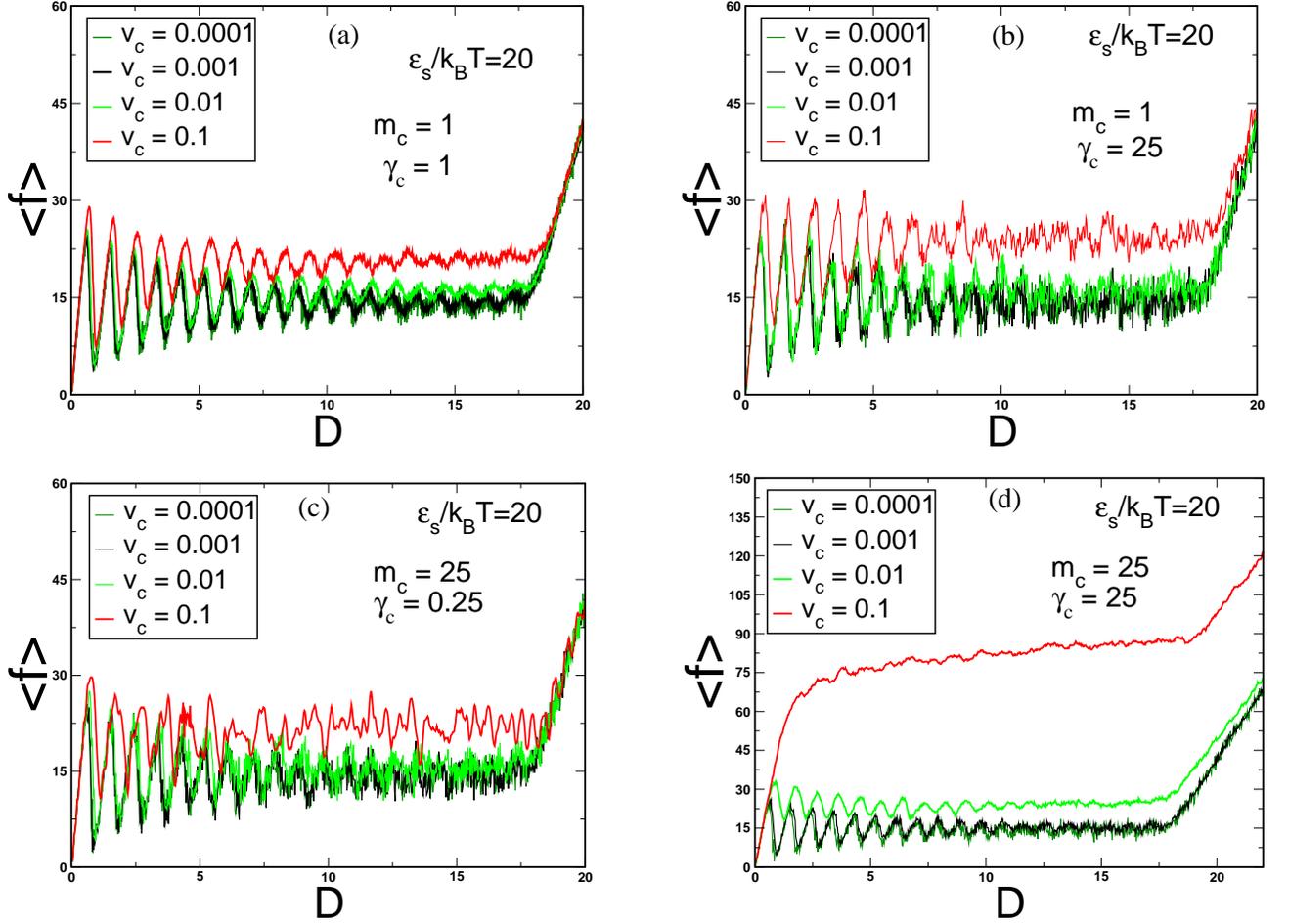

\begin{center}
\vspace{0.3cm}
\includegraphics[scale=0.3]{fvsD_vel.eps}
\hspace{1cm}
\includegraphics[scale=0.3]{fvsD_vel_ml1_gamm25.eps}

\vspace{0.3cm}

\includegraphics[scale=0.3]{fvsD_vel_ml25_gamm025.eps}
\hspace{1.0cm}
\includegraphics[scale=0.3]{fvsD_vel_ml25_gamm25.eps}
\caption{Profiles of  averaged force $\langle f \rangle$ vs distance above the
substrate $D$ displayed for different pulling velocities $v_c$ .
Here $N=20$,  $k_c=50\,\epsilon/\sigma^2$ and $\epsilon_s/k_BT=20$ and (a) $m_c
= 1, \gamma_c = 1$ , (b) $m_c = 1, \gamma_c = 25$, (c) $m_c = 25, \gamma_c =
0.25$, (d) $m_c = 25, \gamma_c = 25$.}
\label{Velocity_Impact}
\end{center}
\end{figure}
Eventually, we demonstrate the impact of cantilever velocity, $v_c$, as well as
of its mass, $m_c$, and friction coefficient, $\gamma_c$, on the force-distance
profile. Apparently, these parameters affect differently strong the observed
force - distance relationship. Similar to the results, obtained for our
coarse-grained model in Sec. \ref{Second_Section}, in the MD-simulation data the
plateau height grows less than twice upon velocity increase of three orders of
magnitude (see Fig.~\ref{Velocity_Impact})! Only for a very massive, ($m_c =
25$), and strong-friction, ($\gamma_c = 25$), cantilever, the plateau height
grows significantly and gains a slight positive slope (see
Fig.\ref{Velocity_Impact}d) whereby oscillations vanish. This occurs for the
fastest detachment $v_c=0.1$. Evidently, this effect is related to the combined
role of the friction force in the case of rapid detachment along with the much
larger inertial force ($m_c=25$) whereby the substrate-induced oscillations are
overshadowed by the increased effort of pulling.  In contrast, neither
Fig.~\ref{Velocity_Impact}b, nor Fig.~\ref{Velocity_Impact}c indicate any major
qualitative changes in the $f$-vs-$D$-behavior when medium-friction, or mass
cantilever alone are drastically changed.

\begin{figure}[ht]
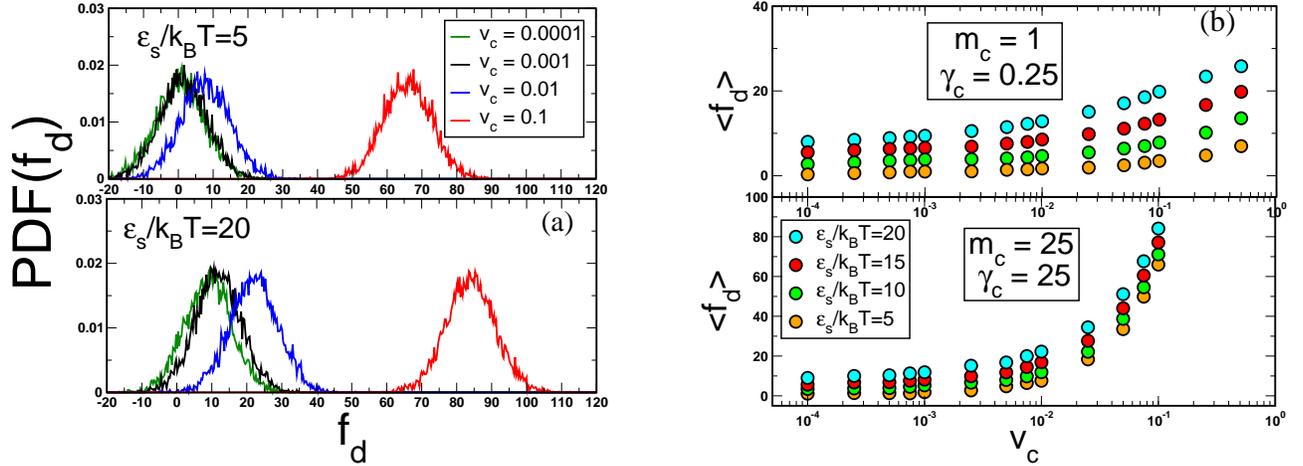

\begin{center}
\includegraphics[scale=0.3]{PDF_forced.eps}
\hspace{1cm}
\includegraphics[scale=0.3]{fdVSv.eps}
\caption{(a) Probability distribution function of a force $f_d$ at the
cantilever measured in the moment of detachment (when the last monomer leaves
the substrate). Here $N=20$, $k_c=50\,\epsilon/\sigma^2$ , $\epsilon_s/k_BT=20$,
$m_c=1$ and $\gamma_c=0.25$. (b) Averaged force  $\langle f_d\rangle$ plotted
versus pulling velocity $v_c$ in the semilogaritmic scale. The values of
parameters are the same. }
\label{PDF_force_MD}
\end{center}
\end{figure}

The PDF of the detachment force and its velocity $v_c$ dependence are shown in
Fig. \ref{PDF_force_MD}. Similarly as in Sec. \ref{Second_Section}, the average
value and dispersion grow with increasing speed of pulling and this is weakly
sensitive with regard to the adsorption strength of the substrate $\epsilon$.
Remarkably, the mean detachment force $\langle f_d\rangle$ shows a similar
nonlinear dependence on $\ln v_c$ (cf. Fig. \ref{Force_Average}a). The
crossover position does not change practically as the adhesion strength is
varied, and the variation of the other parameters ($m_c=1 \to 25,\;
\gamma_c=0.25 \to 25$) towards a massive and strong-friction cantilever
 render this crossover considerably more pronounced.

\begin{figure}[ht]
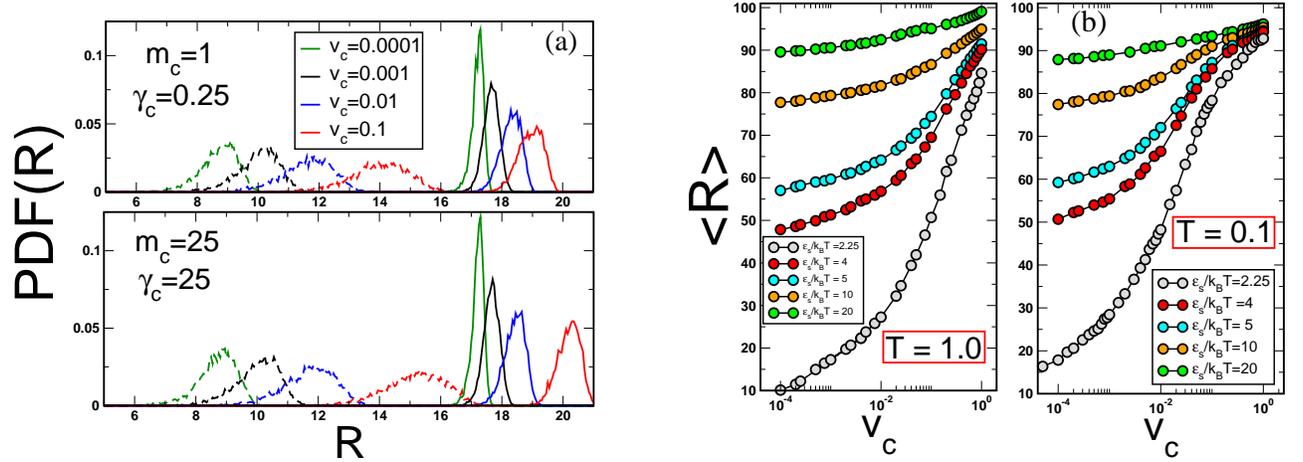

 \includegraphics[scale=0.3]{PDF_R.eps}
\hspace{1cm}
\includegraphics[scale=0.3]{RVSv_N100.eps}
\caption{(a) Probability distribution function of the  detachment height $R$.
Upper panel present results for  $m_c=1$ and $\gamma_c=0.25$ while lower panel
contains data for $m_c=25$ and $\gamma_c=25$.  Solid line represent results for
$\epsilon_s/k_BT=20$ while dashed line for $\epsilon_s/k_BT=5$. Here $N=20$,
$k_c=50\,\epsilon/\sigma^2$. (b) Averaged detachment height $\langle R\rangle$
as a function of pulling velocity $v_c$ in the semi-logarithmic scale plotted
for different adsorption strengths as indicated: (left panel) $k_BT=1.0$, (right
panel) $k_BT=0.1$. On both panels symbols plotted with same color correspond to
the same ratio of $\epsilon_s/k_BT$ while the corresponding values of
$\epsilon_s$ and $k_BT$ differ. Here $N=100$, $k_c=50\,\epsilon/\sigma^2$,
$m_c=1$ and $\gamma_c=0.25$.}
\label{PDF_R_MD}
\end{figure}

The complementary PDF for the detachment height $R$ is given in Fig.
\ref{PDF_R_MD}a together with the corresponding average $\langle R \rangle$
vs $v_c$ relationship. As predicted by our analytic model, cf. Section
\ref{Second_Section}, the height of final detachment of the chain from the
substrate becomes larger for faster peeling $v_c$ and stronger adhesion
$\epsilon$, which is consistent with the MD data. One can see again the typical
sigmoidal-shape in the $\langle R \rangle$ vs $v_c$ dependence.

The two panels for different temperature, shown in Fig. \ref{PDF_R_MD}b,
indicate a smaller increase in $\langle R\rangle$ at the higher temperature,
provided the pulling velocity $v_c$ is sufficiently small too. This can be
readily understood in terms entropic (rubber) elasticity of polymers and
represents a case of delicate interplay between entropy and energy-dominated
behavior. It is well known that a polymer coil becomes less elastic (i.e.,
it contracts) upon a temperature increase, cf. the lowest (grey) curve in Fig.
\ref{PDF_R_MD}b, (left panel) at $T=1.0$, so that $R$ is smaller than in the
corresponding lowest curve for $T=0.1$ in the right panel of Fig.
\ref{PDF_R_MD}b. This occurs at low values of $v_c$. On the other hand, the
softer chain (at $T=0.1$) stretches more easily and, therefore, $R$ goes up to
$\approx 95$ for the highest speed $v_c=10^4$ instead of $R \approx 80$ for
$T=1.0, v_c = 10^4$. This entropic effect is well expressed at weak attraction
to the surface, $\epsilon / k_BT = 2.25$, which does not induce strong
stretching of the bonds along the chain backbone. In contrast, at high $\epsilon
/ k_BT = 20$, the bonds extend so strongly that the chain turns almost into a
string and entropy effects become negligible. The energy cost of stretching
then dominates and leads to higher values of $R$ at the higher temperature (cf.
upper most green symbols in Fig. \ref{PDF_R_MD}b) since it is now the elasticity
of the bonds between neighboring segments which governs the physics of
detachment. In this case the elastic constant of the bonds effectively decreases
with an increase of $T$ so that the distance of detachment $R$ in the left panel
of Fig. \ref{PDF_R_MD}b for $T=1.0$ is higher than that for $T=0.1$ in the right
panel.

\section{Discussion}
\label{sec_results}

We have demonstrated in this paper that a simple theory, based on the Onsager
stochastic equations, yields an adequate description of a typical AFM-
experiment within the displacement-control mode. This approach makes it
possible to
relax most of the restrictions inherent in the BE-model. For example, this
approach also holds for small desorption activation barriers  (i.e., for $E_b
(f) \approx k_B T$), and also for nonlinear barrier vs. force dependence. It
naturally takes into account the reversible desorption-adsorption events
\cite{Diezemann} which are neglected in BE-model. Moreover, it does not rest on
the stationary approximation (which is customary in the standard Kramers rate
calculation \cite{Hanggi}) and is, therefore, ideally suited for description of
driven force- (FC) or displacement-control (DC) regimes. One of the principal
results in this analytic treatment is the predicted existence of characteristic
spikes the mean force vs distance profile, observed in the DC-regime.  These
spikes depend on the adsorption energy $\epsilon$, cantilever
spring constant $k_c$ as well as on the cantilever velocity $v_c$. In
equilibrium, this has been found earlier by Jagota and coworkers
\cite{Jagota_1}. The PDF  of detachment forces and detachment distances are been
thoroughly investigated. The relevant mean detachment force is found to be a
strongly nonlinear function of $v_c$ which is mainly governed by the  nonlinear
chain stretching upon increasing $v_c$. The average full detachment (peeling)
time scales $\propto 1/v_c$ which is supported by early theoretical findings
\cite{Seifert}.

Some of these predictions were checked by means of MD-simulation and found in a
qualitative agreement with the results, gained  by the analytic method. Most
notably, this applies to properties like the characteristic force oscillations
pattern and the mean force vs cantilever velocity $v_c$ dependence. On the other
hand, our MD-simulation reveals a very strong  increase in the magnitude of the
force plateau for a strong-friction ($\gamma_c = 25$) and massive ($m_c = 25$)
cantilever. Interestingly, in this case the spikes pattern is almost totally
smeared out. This might be the reason why the force spikes pattern is not seen
in laboratory detachment experiments. We recall that in a recent Brownian
dynamic simulation (which totally ignores inertia forces) \cite{Jagota_3}, the
friction coefficient of the cantilever was $70$ times larger than the friction
coefficient of the chain segments. It was shown that for this high-friction
cantilever and large velocity of pulling, the force spikes pattern was
significantly attenuated \cite{Jagota_3} so that information on the base
sequence was hardly assessable. Therefore, fabrication of a stiff and
super-light, nanometer-sized AFM probe would be a challenging task for future
developments of biopolymer sequencing.

As an outlook, our coarse-grained Onsager stochastic model could be
generalized
to encompass investigations of forced unfolding of a multi-domain,
self-associating  biopolymers \cite{Kreuzer_4}.

\subsection*{Acknowledgement}

This work was supported by grant SFB 625 from German Research Foundation (DFG).
Computational time on PL-Grid Infrastructure is acknowledged.
A.M. thanks the Max Planck Institute for Polymer Research in Mainz, Germany,
and CECAM - MPI for hospitality during his visit at the Institute.

\begin{appendix}
 \section{The separation $R$ as an instantaneously adjustable variable}
\label{App}

Due to strong adsorption, the desorbed portion of polymer chain is expected to
be strongly stretched. One could simplify the force, $f_{R} \approx (R/bn) [3 -
(R/bn)^2]/[1 - (R/bn)^2] \approx 1/[1 - (R/bn)]$, where Eq.
(\ref{Interpolation}) has been used and the contribution of the cantilever has
been neglected. Therefore, the simplified equation which governs $R$ reads
\begin{eqnarray}
 \dfrac{d R}{d t} \approx - \dfrac{\lambda_{R} k_BT}{b [1 - (R/bn)]}
\label{Simplified}
\end{eqnarray}
This equation can be easily solved and the corresponding solution has the form
\begin{eqnarray}
 \dfrac{1 - R(t)/bn}{1 - R_0/bn} = \sqrt{1 + t/\tau_{R}}
\end{eqnarray}
where the relaxation time $\tau_{R} = (b^2/2k_BT\lambda_{R}) (1 - R_0/bn)^2$.
This result suggests that for a strongly stretched chain, i.e., for
$R_0\leqslant bn$,  the relaxation time $\tau_{R}$ is very small \cite{Febbo}.
For example, in the case that $R_0 = b (n - 1)$  we have
\begin{eqnarray}
\tau_{R} = \dfrac{
b^2}{2k_BT\lambda_R n^2}
\label{Tau_R}
\end{eqnarray}

This relaxation time should be compared to the characteristic time, $\tau_{\rm
Kram}$, of the slow variable $n(t)$ which is governed by the Kramers process.
According to the semi-phenomenological Bell model \cite{Bell}, the
characteristic time of unbonding (that is, desorption in our case) is given by
$\tau_{\rm Kram} = \tau_0 \exp [(\Delta E - r_0 f_p)/k_BT]$ where $\tau_0 =
\xi_0 b^2/k_BT$ is the segmental time, $\Delta E = F_1 - F_2$ is the activation
energy for single monomer desorption, $r_0$ stands for the width of adsorption
potential, and $f_p$ is the plateau height. The free energies in the desorbed,
$F_1$, and in the adsorbed, $F_2$, states are given by $F_1 = - k_BT \ln \mu_2$
and $F_2 = - k_BT \epsilon - k_BT \ln \mu_3$ where $\mu_2$ and $\mu_3$ are the
so-called connective constants in two- and three dimensions respectively
\cite{Vanderzande}.

As mentioned in Sec. \ref{MM}, for large adsorption energies $\epsilon$ the
dimensionless plateau height ${\widetilde f}_p \stackrel{\rm def}{=} b f_p/k_BT
\propto \epsilon$. Taking this into account, one could represent $\tau_{\rm
Kram} $ in the following form:
\begin{eqnarray}
 \tau_{\rm Kram}  = \dfrac{\tau_0 \mu_2}{\mu_3} \: \exp[(1 - \alpha) \epsilon],
\label{KK}
\end{eqnarray}
where $\alpha = r_0/b < 1$. Therefore, in the case when  $\tau_R \ll \tau_{\rm
Kram}$, the distance $R$ could be treated as the fast variable. With Eqs.
(\ref{Tau_R}) and (\ref{KK}) and taking into account that the  Onsager
coefficient  $\lambda_R = 1/\xi_0 n$, this condition means that
\begin{eqnarray}
 \dfrac{n \mu_2}{\mu_3} \: \exp[(1 - \alpha) \epsilon] \gg 1.
\label{gg}
\end{eqnarray}
This condition holds for all typical values of the relevant parameters.

\section{Runge-Kutta algorithm for stochastic differential-algebraic equations}
\label{R-K}

In order to solve the DAE (\ref{DAE}) numerically, one may employ the second
order Runge-Kutta (RK) algorithm. To this end the first equation in
Eq.~(\ref{DAE}) may be rewritten as an integral equation which relates the
$i$-th and $i+1$ grid points (using discrete time points $t_i = i h$)
\begin{eqnarray}
 n_{i+1} = n_{i} + \lambda_n \int\limits_{t_i}^{t_{i+1}} \: f_n (n(s), R(s)) ds
+ w_n (h)
\label{Integral_Equation_2}
\end{eqnarray}
where $h$ is the time step and $n_i = n(i h)$. Moreover, $w_{n} (h)
=\int_{t_i}^{t_{i+1}} \; \xi_n (s) ds$ describes a Wiener process with zero
mean and with variance:
\begin{eqnarray}
 \langle w_{n} (h) w_{n} (h)\rangle &=& \int_{t_i}^{t_{i+1}} d s_1 \;
\int_{t_i}^{t_{i+1}} d s_2 \langle  \xi_n (s_1) \xi_n(s_2)\rangle =
2 \lambda_n k_BT \: \int_{t_i}^{t_{i+1}} d s_1
\;\int_{t_i}^{t_{i+1}} d
s_2 \; \delta (s_1 - s_2) \nonumber\\
&=& 2 \lambda_n k_BT \: h
\end{eqnarray}

The  integral over the deterministic force in Eq. (\ref{Integral_Equation_2})
within this $2-nd$ order approximation reads
\begin{eqnarray}
 \int_{t_i}^{t_{i+1}} \: f_{n} (n(s), R(s)) \: ds \approx
\dfrac{h}{2} \left[ f_{n} (n_{i+1}, R_{i+1}) +  f_{n}
(n_{i}, R_{i}) \right] + {\cal O} (h^3)
\end{eqnarray}
This  is so-called trapezoidal rule for approximation of the integral. In order
to calculate $f_{n} (n_i, R_i)$, one  should first take the initial value $n_i$,
and find $R_i$ through the solution of the nonlinear equation $G(n_i, R_i, t_i)
= 0$. For the calculation of $f_{n} (n_{i+1}, R_{i+1})$, one can use the
forward Euler method of order $1$, i.e., $n_{i+1}^E = n_i + h \lambda_n f_{n}
(n_i, R_i) + h^{1/2} (2 \lambda_n k_BT)^{1/2} Z_n$ and $R_{i+1}^E$ are obtained
as solution of the equation $G(n_{i+1}^E, R_{i+1}^E, t_{i+1}) = 0$. Here the
random variable $Z_n$ is Gaussian with zero mean value and with variance
\begin{eqnarray}
 \langle Z_n^2 \rangle = 1
\label{Z_2}
\end{eqnarray}

As a result, the recursive procedure which relates the $i$-th and $i+1$ grid
points can be defined as:
\begin{enumerate}
 \item For a given initial value of $n_i$, go to Eq. (\ref{G}) or  Eq. (\ref{P})
and solve this nonlinear equation (e.g. $G(n_i, R_i, t_i) = 0$) with respect to
$R_i$.
\item Compute $g_1 = f_{n} (n_i, R_i)$.
\item Compute $n_{i+1}$ and $R_{i+1}$ within the Euler approximation, i.e.,
calculate first $n_{i+1}^E = n_i + h \lambda_n f_{n} (n_i, R_i) + h^{1/2} (2
\lambda_n k_BT)^{1/2} Z_n$ and then solve $G(n_{i+1}^E, R_{i+1}^E, t_{i+1}) = 0$
with respect to $R_{i+1}^E$.
\item  Compute $g_2 = f_{n} (n_{i+1}^E, R_{i+1}^E)$.
\item Compute the corrected  $n_{i+1}$ , i.e.
\begin{eqnarray}
n_{i+1} = n_i  + \dfrac{h}{2} (g_1 + g_2) +  h^{1/2} (2 \lambda_n
k_BT)^{1/2} Z_n \nonumber
\end{eqnarray}
\item Finally, with the value of $n_{i+1}$, go to item $1$  and  solve the
nonlinear equation $G(n_{i+1}, R_{i+1}, t_{i+1}) = 0$ with respect to $R_{i+1}$.
\end{enumerate}

\end{appendix}

\end{document}